\documentclass[aps,prb,reprint,groupedaddress]{revtex4-1}
\usepackage[pdftex]{graphicx,hyperref}
\usepackage{dcolumn,physics}
\usepackage{bm,comment}
\usepackage{color}

\begin{document}
\preprint{}
\title{Dimensional reduction in quantum spin-1/2 system on a 1/7-depleted triangular lattice}
\author{Ryo Makuta}
  \email{makuta-r@g.ecc.u-tokyo.ac.jp}
\author{Chisa Hotta}%
   \email{chisa@phys.c.u-tokyo.ac.jp}
     \affiliation{Department of Basic Science, University of Tokyo, 3-8-1 Komaba, Meguro, Tokyo 153-8902, Japan}
\date{\today}
\begin{abstract}
We study the magnetism of a quantum spin-1/2 antiferromagnet on a maple-leaf lattice
which is obtained by regularly depleting 1/7 of the sites of a triangular lattice.
Although the interactions are set to be spatially isotropic,
the ground state shows a stripe N\'eel order and the temperature dependence of magnetic susceptibility
follows that of the one-dimensional XXZ model with a finite spin gap.
We examine the nature of frustration by mapping the low energy degenerate manifold of states to the
fully packed loop-string model on a dual cluster-depleted honeycomb lattice,
finding that the order-by-disorder due to quantum fluctuation characteristic of highly frustrated magnets
is responsible for the emergent stripes.
The excited magnons split into two spinons and propagate in the one-dimensional direction along the stripe,
which is reminiscent of the XXZ or Ising model in one dimension.
Unlike most of the previously studied dimensional reduction effects,
our case is purely spontaneous as the interactions of the Hamiltonian
retains a two-dimensional structure.
\end{abstract}
\maketitle

\section{Introduction}
\label{sec1}
Highly frustrated magnets exhibit nearly degenerate low-energy states
as a consequence of competition between different local interactions that cannot be simultaneously satisfied.
For such cases, the standard magnetic orderings are strongly suppressed
and the system either remains disordered down to the lowest temperature
or experiences exquisite sensitivity to degeneracy-breaking perturbations.
Once a large degeneracy is resolved, some unexpected orders may emerge,
which is commonly referred to as ``order-by-disorder"\cite{Villain1980}.
Whether the order-by-disorder is achieved or not depends
on the degree of frustration which is often determined by a lattice geometry.
A classical example is a face-centered-cubic vector antiferromagnet
which shows collinear or noncollinear orderings due to perturbations\cite{Henley1987,Gvozdikova2005}.
Whereas, in a classical pyrochlore antiferromagnet an extremely strong frustration
does not even allow for an order-by-disorder and the system keeps a highly disordered character referred to as spin-ice\cite{Ramirez1999,Bramwell2001}.
In quantum magnets, quantum fluctuations may play a role in degeneracy-breaking perturbation,
and via order-by-disorder, a supersolid phase appears in a triangular lattice XXZ model\cite{Heidarian2005,Wessel2005,Melko2005}.
The Heisenberg triangular lattice antiferromagnet with larger quantum fluctuation
turned out to have a 120$^\circ$ N\'eel ordered ground state\cite{Jolicoeur1989,Bernu1994,White2007},
although in the early stage an interplay of large quantum fluctuation and large frustration
is expected to produce a resonating valence bond liquid\cite{Fazekas1974,Anderson1987}.
Finally, for a more highly frustrated kagome lattice,
a quantum mechanical disordered spin liquid is realized at zero temperature
\cite{liao2017,depenbrock2012,yan2011,nishimoto2013}.
\par
Although the concept of geometrical frustration and the order-by-disorder have existed for years
and had been a source of abundant magnetic and nonmagnetic phases of matter,
we still lack enough clues to understand the degree of frustration and to control them.
This shall be because the platform is limited to a few lattice structures including kagome, pyrochlore,
triangular, checkerboard lattices and their variants.
%
\begin{figure}[b]
  \includegraphics[width=0.5\textwidth]{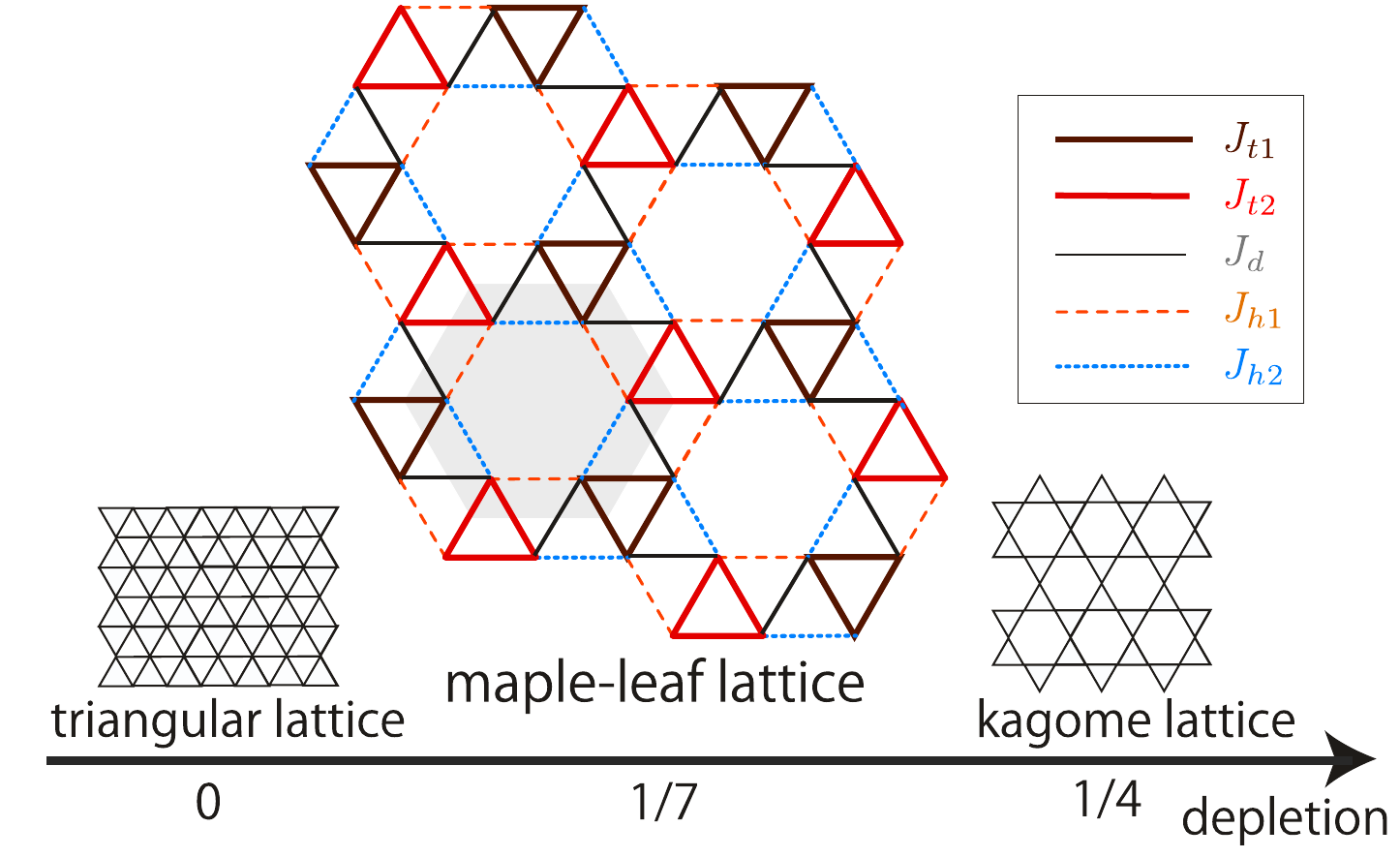}
  \caption{The relationship between triangular lattice, maple-leaf lattice,
   and kagome lattice and definition of exchange interaction constants.
   Five exchange constants of a maple-leaf lattice and the unit cell (shaded) including six sites are shown. }
\label{f1}
\end{figure}
\par
The geometry of the above mentioned frustrated two-dimensional lattices are interrelated;
the kagome lattice can be realized by regularly depleting 1/4 of the sites of the triangular lattice.
The effect of gradually weakening the interactions of the depleted sites with their surroundings
is examined both for the ground state and for the thermodynamic quantities\cite{Arrachea2003,Koretsune2009}.
The finite temperature double-peak specific heat and the variance of susceptibility
obtained by the exact diagonalization (ED) on a small cluster
indicates that the low energy properties of the two models may be smoothly interpolated\cite{Koretsune2009}.
However, a more precise analysis of the excitation spectra showed that
even though the N\'eel order is suppressed when the interaction ratio
between the depleted and the regular bonds is less than 1/5,
the full depletion limit of the triangular lattice antiferromagnet does
not continue to the kagome lattice ones\cite{Arrachea2003}.
These works indicate that the frustration effect depends very sensitively on the geometry of the lattice
and is not easy to understand.
\par
Here, we find another order-by-disorder phenomenon, a dimensional reduction effect,
in a maple-leaf lattice antiferromagnet.
The reduction of three-dimensionality of layered systems to effectively two-dimensions
is previously reported both in theories and in experiments\cite{Sebastian2006,Roesch2007,Grando2013,Okuma2021},
which is attributed to the competing frustrated inter-layer interactions.
In two dimensional systems, a one-dimensional spinon-like continuum excitation is observed
in a metal-organic square lattice antiferromagnet\cite{Skoulatos2017}
as well as in the triangular lattice magnet\cite{Coldea2003,Kohno2007}.
The spinon excitation in two-dimensional antiferromagnets
shares the same concept as fractional charges\cite{Hotta2008} or equivalently,
fractons in frustrated electronic systems.
However, in these systems, the dimensional reduction is encoded in the
Hamiltonian; it is induced by weakening the strength of frustrated interaction associated with the reduced dimensionality, e.g. inter-layer coupling or inter-chain coupling.
By contrast, in the present system, the spontaneous reduction of dimensionality occurs for a purely
two-dimensional geometry of interactions.
We discuss this context more precisely in \S.\ref{sec:summary}.
\par
A maple-leaf lattice is a family of geometrically frustrated lattices based on a triangular unit as shown
in Fig.\ref{f1}.
It can be obtained by periodically depleting 1/7 of the sites of a triangular lattice,
and is located in the diagram in between the triangular and kagome lattices.
Experimental realizations of maple-leaf structure are reported
both in natural minerals\cite{Fennell2011,kampf2013,mills2014}
and in man-made crystals\cite{cave2006,aliev2012,Haraguchi2018,Bluebellite}.
In theories, the ED study shows that
the ground states of Heisenberg antiferromagnet on the maple-leaf lattice
with spatially uniform interactions hosts six-sublattice long-range order similar to the
one found in the classical counterpart\cite{Schulenburg2000,Schmalfuss2002}.
However, this long-range order may disappear when one chooses $J_d$ in Fig.~\ref{f1} from a uniform $J$
and increase them up to $J_d/J>1.45$\cite{Farnell2011}.
These results indicate that this lattice provides another rich platform
to make a comparative study of the role of geometry and the degree of frustration.
\par
Recently, a new family member of the maple-leaf lattice called
bluebellite (Cu$_6$I$_6$O$_3$(OH)$_{10}$Cl) is found to show a particular magnetic susceptibility
that almost perfectly matches the Bonner-Fisher curve\cite{Bonner1964} of
a purely one-dimensional $S=1/2$ Heisenberg antiferromagnet\cite{Bluebellite}.
A similar Bonner-Fisher-like magnetic susceptibility is observed
in a $S=3/2$ maple-leaf lattice antiferromagnet, Na$_2$Mn$_3$O$_7$\cite{Venkatesh2020}.
These results may suggest that there is an inherent nature in maple-leaf lattice
that spontaneously reduces the dimensionality.
\par
In this paper, we study the interplay of frustration and quantum fluctuation
for this 1/7-depleted triangular lattice.
In \S.\ref{sec2} we consider an antiferromagnetic Ising model and clarify the nature and the degree of frustration of the lattice.
In \S.\ref{sec3} we perform an ED study and show that the ground state is a symmetry-broken stripe phase.
In \S.\ref{sec4} we examine a temperature dependence of susceptibility
by varying the interaction parameters.
We find that the low energy magnetic excitation shows a similar feature to that of
the one-dimensional spin-1/2 XXZ model having a spin-gapped antiferromagnetically ordered ground state.
By further introducing an inter-layer coupling to our model,
the temperature-dependent profile of our susceptibility is modified to the one that
resembles the Bonner-Fisher curve, except at a very low temperature region
where our susceptibility shows an exponential decrease due to the spin-gap.

\section{Model}
\label{sec2}
\subsection{Maple-leaf lattice Heisenberg model}
\par
We consider a spin-1/2 Heisenberg model on the maple-leaf lattice with nearest neighbor interactions
whose Hamiltonian is given as
\begin{equation}
  \mathcal{H}=\sum_{\langle i, j\rangle}J_{i,j}\bm{S}_i\cdot\bm{S}_j,
  \label{Heisenberg}
\end{equation}
where $\bm{S}_i$ is a spin-1/2 operator on site $i$
and the summation is taken over all $\langle i, j\rangle$ pairs of neighboring sites.
The Heisenberg model is the isotropic limit of the XXZ model given as
\begin{equation}
  \mathcal{H}=\sum_{\langle i, j\rangle}
  J^{xy}_{i,j}(S^{x}_{i}S^{x}_{j}+S^{y}_{i}S^{y}_{j})+J^{z}_{i,j}S^{z}_{i}S^{z}_{j},
  \label{Heisenberg2}
\end{equation}
where $S^{\alpha}_i$ is the $\alpha=x,y$ and $z$ component of spin,
and the $J^{xy}_{i,j}=J^{z}_{i,j}$ case corresponds to Eq.(\ref{Heisenberg}).
In the following, the $J^{xy}_{i,j}$- and $J^{z}_{i,j}$-terms are often separately examined.
Since this maple-leaf lattice belongs to space group $R3$,
the Heisenberg spin exchange interaction $J_{i,j}$ consists of five species,
$J_{t1},\,J_{t2},\,J_d,\,J_{h1},\,J_{h2}$, whose spatial arrangement is shown in Fig. \ref{f1}(a).
Experimentally, in a bluebellite the structural analysis and the information on the alignment of $d$-orbials which carry spin-1/2
suggests that the sign and amplitude of these interactions are $J_{t2} \geq J_{d}\geq J_{t1} \geq J_{h1}\geq 0 \geq J_{h2}$.
Whereas, for simplicity and to clarify the intrinsic nature of the geometry of lattice,
we set $J_{t1}=J_{t2}=J_{d}=J,$ $J_{h1}=-J_{h2}=J_h$ and vary $J_h/J$ as a parameter
in the main part of the calculation.
We denote the total number of sites as $N$ and the number of unit cells as $N_{\rm cell}$,
where $N=6N_{\rm cell}$ for the maple-leaf lattice and $N=7N_{\rm cell}$ for the corresponding triangular lattice.

\subsection{Ising limit}
We first examine the nature of frustration of the maple-leaf lattice through the comparison with the triangular lattice.
To this end, consider an Ising model
which is the classical limit of the Heisenberg model,
where the quantum fluctuation due to spin exchange is neglected
by taking $J^{xy}_{i,j}=0$ in Eq.(\ref{Heisenberg2}).
\par
Let us remind that the ground state of an antiferromagnetic Ising model on a uniform triangular lattice with $J^z_{ij} =J >0$
consists of massive numbers of states which altogether contribute to the residual entropy amounting
to $S=0.323k_{\mathrm{B}}N$\cite{Wannier1950}.
The local constraint of Ising spins that belong to this degenerate ground state
is to have either two-up one-down spins or two-down one-up spin on a triangle;
the state is called UUD/DDU, whose Ising energy is $E_{\rm ising}= -NJ/4$.
\par
\begin{figure}
  \includegraphics[width=0.50\textwidth]{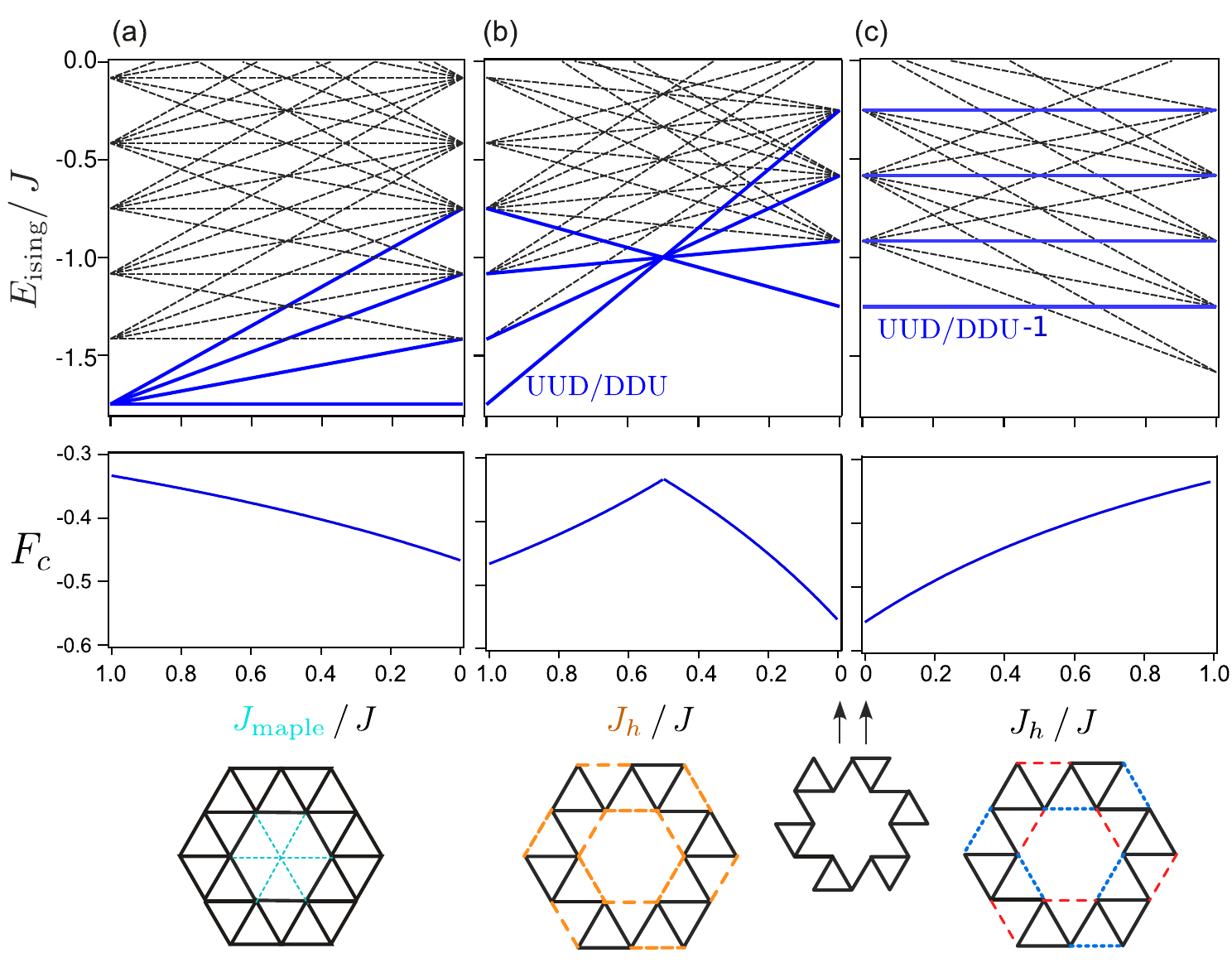}
  \caption{Ising energy $E_{\rm ising}=\langle {\cal H}_{\rm ising} \rangle/N_{\rm cell}$
  per unit cell for the lowest few levels in the $S_z^{(\mathrm{tot})}=0$ sector
  obtained for 18 site cluster by varying the geometry of the lattice.
  Here, we introduce $J_{\rm maple}$ which are the six bonds inside the hexagon
  depleted in constructing the maple-leaf structure from a triangular lattice.
  (a) $J_{\rm maple}$ varied by fixing $J=J_{t1}=J_{t2}=J_d=1$ (antiferromagnetic),
  (b) $1\ge J_{h1}=J_{h2}\ge 0$ (antiferromagnetic) varied by taking $J_{\rm maple}=0$,
  where $J_{h1}=J_{h2}=0$ on the r.h.s. has the structure consisting of triangles
  connected with each other by a single bond $J_d$.
  (c) $J_{h1}=-J_{h2}\ge 0$ around the hexagon consisting of antiferromagnetic and ferromagnetic bonds
   is varied which is the structure we mainly focus on.
   For all panels, the solid lines correspond to the UUD/DDU states and the broken lines are
   the states with UUD/DDU partially violated.
   The constraint parameter $F_c$ is examined, where $F_c=-1$ corresponds to the nonfrustrated case
   and $F_c \sim 0$ to the highly frustrated limit. }
  \label{f2}
\end{figure}

To see how the lattice geometry changes the low energy structure of the model,
we gradually vary the strength of $J_{\rm maple}$, which are the six bonds inside the hexagon to be depleted
where $J_{\rm maple}=0$ corresponds to the full 1/7-depletion of the triangular lattice.
As shown in Fig.~\ref{f2}(a) the UUD/DDU states split into several levels with equal energy spacings,
$E_{\rm ising}=-J(N/4-n)-J_{\rm maple}n$, where $n=0, 1, ..., N/7$,
while the energy of the lowest level remains unchanged.
\par
For a 1/7-depleted structure at $J_{\rm maple}=0$ we can further decrease the strength of bonds around
the depleted hexagon as $J_{h1}=J_{h2}=1$ to 0, as shown in Fig.~\ref{f2}(b),
where all the UUD/DDU levels vary linearly
as $E_{\rm ising}=-J(N/24+n)+J_h (2n-N/4)$ where $n=0, 1, ..., N/6$,
and all the UUD levels cross at $J_{h1}=J_{h2}=0.5J$ which has a comparably large degeneracy
with the original triangular lattice.
Finally, in Fig.~\ref{f2}(c) we introduce the alternating ferromagnetic and antiferromagnetic bonds along the hexagon
and increase their amplitude, finding that all the UUD/DDU levels remain unchanged.
The higher energy levels shown in broken lines are the non-UUD states,
namely their UUD/DDU structure is partially broken. With increasing $\pm J_h$,
they descend and overtake the perfect UUD/DDU at $J_{h1}=-J_{h2}\ge 0.5$.
The parameter used in Fig.~\ref{f2}(c) is the one we mainly focus on.
\par
The frustration generated by the competition between different exchange interactions
usually leads to large classical ground-state degeneracies.
In the present case,
the upper bound of the degeneracy of the lowest energy UUD/DDU manifold
for $J_{h1}=J_{h2}(>0.5)$ is roughly evaluated as
$2^{N/6}$ with a corresponding residual entropy of $S=0.116k_{\rm B}N$,
and for $J_{h1}=-J_{h2}$ as $6^{N/6}$ with $S=0.299k_{\rm B}N$,
which are explained as follows.
In the former antiferromagnetic $J_h$,
the lowest energy UUD/DDU states satisfy the condition that
the spins align in a staggard manner along the hexagons to maximally gain $J_h$.
For each hexagon, one can prepare two such states, which give $2^{N/6}$.
However, among them, those giving UUU or DDD configurations
to $J_{t1}$ or $J_{t2}$ triangles should be excluded.
For three adjacent hexagons, 2 states among $2^3$ are excluded,
while for four adjacent hexagons, again 2 states among $2^4$ are excluded,
which means that the lower bound of degeneracy is still as large as $2^{N/6}(3/4)^{N/6}$
leading to the massive residual entropy.
\par
In the latter case, the local constraint for the lowest energy UUD/DDU states is to have
both sides of spins on $J_d$ bond align antiparallel.
Therefore, once the UUD/DDU spin configurations on $N/6$ different $J_{t1}$-triangles
are determined, the spin configurations on all $J_{t2}$ triangles, namely the whole rest
of the spins are automatically determined through $J_d$ bonds.
Among them, those giving UUU/DDD to $J_{t2}$ triangle should be excluded.
For three adjacent $J_{t1}$ triangles, among $6^3$ UUD/DDU states the ones that give
UUU/DDD on $J_{t2}$ enclosed by these three is 54.
Therefore, the lower bound of degeneracy is $6^{N/6} (162/216)^{N/6}$.
Notice that for both cases, the lowest UUD/DDU manifold includes the state outside the UUD/DDU manifold of
the triangular lattice, and the former is not the subspace of the latter, as we discuss shortly.
\par
Another quantitative measure of frustration given by Lacorre \cite{Lacorre87}
is the constraint function $F_c= -E_0/E_b$
where $E_0$ is the ground state energy and $E_b= -\sum_{\langle i,j\rangle} |J^{z}_{ij}| (S_i^z S_j^z)_{\rm max}$.
When $F_c=-1$ all the local bond energies are optimized simultaneously and
there is no frustration, whereas $F_c \sim 0$ means that the competition
of local interaction leads practically to no energy gain.
As shown in the lower panels of Fig.~\ref{f2}(a)-(c), $F_c$ first decreases and becomes
less frustrated in depleting $J_{\rm maple}$.
With tuning the sign and values of $J_h$,
it becomes as frustrated as the original triangular lattice Ising model,
e.g. when $J_{h1}=-J_{h2}=J$ or $J_{h1}=J_{h2}=J/2$.
\begin{figure*}
  \includegraphics[width=1.0\textwidth]{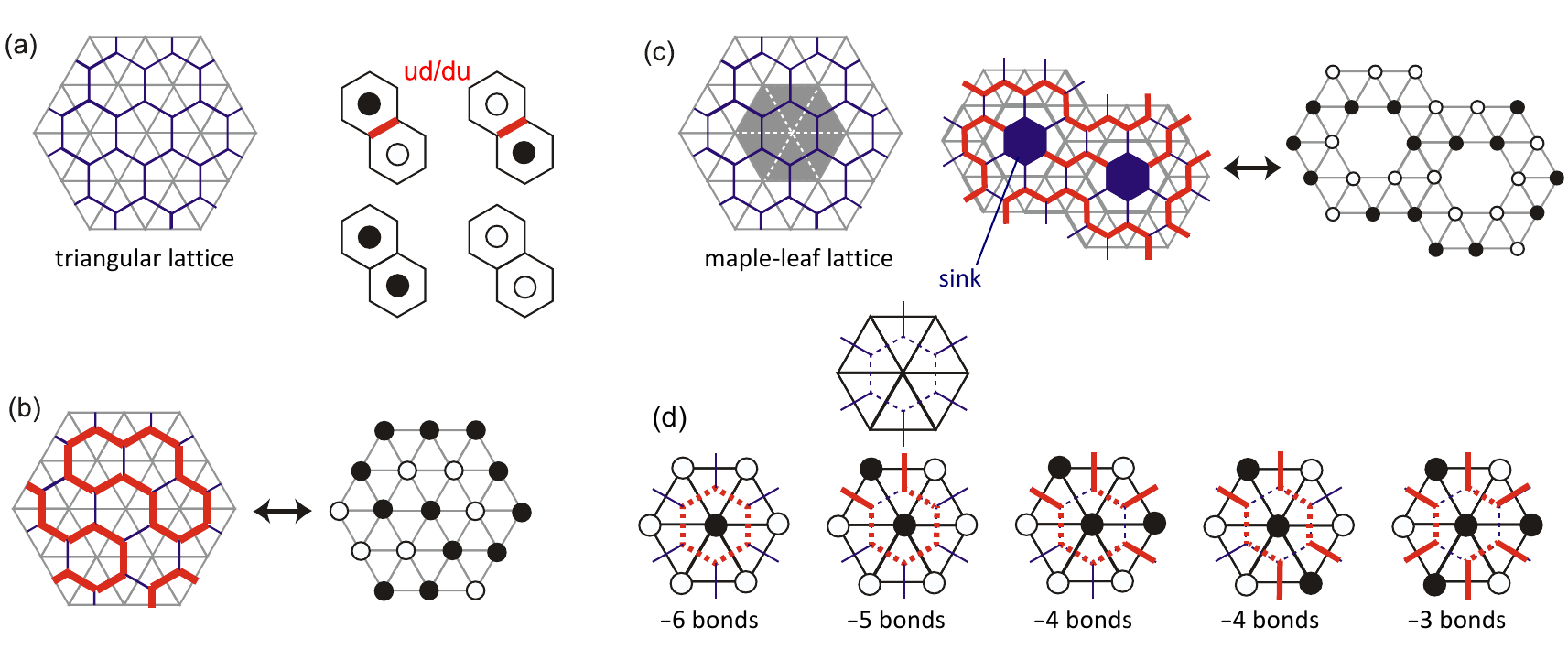}
  \caption{
  (a) Relationship between the triangular lattice (gray lines) and its dual lattice (blue lines).
      Red bold bonds are placed on the honeycomb bond which crosses the up-down pairs of spins on the triangular lattice.
  (b) One of the examples of the UUD/DDU structures on a triangular lattice
      described by the fully packed loop model on a honeycomb lattice.
  (c) Correspondence of the maple-leaf lattice and its dual lattice, a honeycomb lattice with sinks.
   The two right panels are the example of the loop-string model on the honeycomb lattice with sinks
   and the corresponding UUD state of the maple-leaf lattice.
  (d) All the UUD/DDU patterns on the hexagonal unit around the depleted site.
      Broken bonds are part of the loops to be depleted, where the number of these bonds distribute between -6 to -3. }
  \label{f3}
\end{figure*}
\par
To give an intuitive understanding of the above mentioned behavior of low energy levels,
we introduce a dual lattice description of the model.
Figure~\ref{f3}(a) shows a honeycomb lattice which is a dual lattice of the triangular lattice.
One can map the UUD/DDU states of a triangular lattice
to the fully packed loop-covering states on a honeycomb lattice in the following manner;
when the neighboring two spins on a triangular lattice is antiparallel,
the edge of the dual lattice in between these spins is filled by a bold red bond.
Since each triangle has two antiparallel pairs of spins, there is a local constraint
that each honeycomb site is connected to two red bonds,
and resultantly this red bond always forms a closed-loop while never cross with each other
(see Fig.~\ref{f3}(b)).
By counting the total length of the loops the Ising energy is obtained,
which takes a maximum, i.e. $2N$, for the UUD state.
A fully-packed loop model is spanned by the restricted Hilbert space consisting only of
these loop states and serves as a low energy effective model of
strongly correlated quantum spin and charge systems on a triangular lattice and kagome lattices
\cite{Karlo14}.
\par
Depleting 1/7-sites from the triangular lattice modifies the nature of loops on its dual lattice.
Figure~\ref{f3}(c) shows the dual lattice of the maple-leaf lattice,
where the honeycomb lattice is modified such that the hexagons of the honeycomb lattice
surrounding the depleted triangular sites are erased.
We call this vacant unit hexagon a ``sink".
Since the center triangular site is depleted, the hexagonal bonds surrounding the sink are never filled by loops or bonds.
Then, some of the fully packed loops of a honeycomb lattice that crossed
the sink can no longer form a closed loop, and their open edges enter the sink.
We call this description a fully packed loop-string state,
whose example is shown in the right panel of Fig. \ref{f3}(c).
Due to this modification, the total length of the loops and strings are not necessarily
the same although they all describe the UUD/DDU state of the maple-leaf lattice.
Namely, the length of the loop and string varies according to the number of loop-bonds
that belonged to the sink and are lost by depletion.
This splits the originally degenerate ground-state manifold of the triangular lattice model
into equally spaced hierarchial energy levels (see Fig.~\ref{f2}).
As shown in Fig.~\ref{f3}(d), the number of depleted red broken bonds is between 3 to 6.
Each depleted bond carries $0.25J_{\rm maple}$ and the energy level spacing amounts to $0.5J_{\rm maple}$
per six-site unit cell(see Fig.~\ref{f1}).

\begin{figure}
  \includegraphics[width=0.5\textwidth]{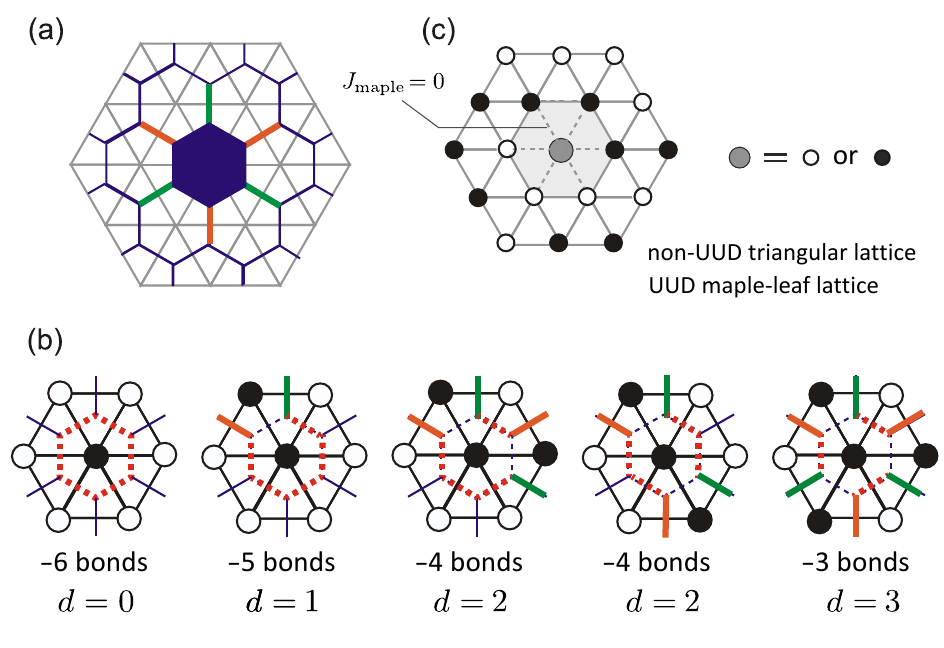}
  \caption{(a) Assigning two different colors to the edges leading to the sinks.
   Red and green lines denote the antiferromagnetic $J_h$ and ferromagnetic $-J_h$.
   (b) Five UUD/DDU patterns on the hexagonal unit around the depleted site with $d=0,1,2,3$,
    where $2d$ is the number of edges of loops that enter the sink.
    Broken lines are part of the loops inside the sink that disappear due to depletion, which amount to $6-d$.
  (c) UUD/DDU states on the maple-leaf lattice which were originally a non-UUD/DDU states of
   the triangular lattice. Inside the hexagon, we have UUU or DDD triangles.
}
  \label{f4}
\end{figure}

In a maple-leaf lattice, the Ising energy further changes by varying
the bond strength along the hexagon $J_{h1}=\pm J_{h_2}>0$
while keeping $J_{t1}=J_{t2}=J_d=J$.
For each sink, even number of strings enter which we denote as $2d$ with $d=0,1,2,3$.
The bonds lost by depletion are shown in broken lines in Fig.~\ref{f3}(d).
Their number is given as $6-d$, and the $2d$ bonds carry $\pm J_h/4$ instead of $-J/4$.
When all the bonds are antiferromagnetic as $J_{h2}=J_h>0$,
the Ising energy for each level in Fig.~\ref{f2}(b) measured from
the classical Ising energy on the triangular lattice is given as
$E_{\rm ising}=\sum_i (6-2d_i)J/4 + (-J_h+J)(d_i-3/2)$
and depend on $J_h$.
However, when half of the bonds on the hexagon is ferromagnetic as $J_{h2}=-J_h<0$,
the contribution from the hexagons are canceled out and we find a $J_h$ independent
profile of energy
$E_{\rm ising}=\sum_i (6-2d_i)J/4+J(d_i-3/2)$ in Fig.~\ref{f2}(c);
as shown in Fig.~\ref{f4}(a) we need to assign different colors to six bonds that
may cross the edges of the sink, which contribute to the Ising energy as $\pm J_h$.
Then, for all five different configurations of strings entering the sink
shown in Fig.~\ref{f4}(b) the number of green and red colored bonds are always
equal, which is the reason for the cancellation.
\par
We finally mention that there are other non-negligible states
that belong to the UUD/DDU states of the maple-leaf lattice but
do not continue to the UUD/DDU states of the triangular lattice.
One example is shown in Fig.~\ref{f4}(c).
The three up spins and three down spins form neighbors around the hexagon of a triangular lattice
whose center spin is to be depleted.
In this case, two triangles inside the hexagon are either UUU or DDD, regardless of the
orientation of the center spins.
However, these triangles are wiped out by the depletion.
Therefore, the UUD states of the maple-leaf lattice are not the subspace of the UUD states of the triangular lattice.
We show in Appendix~\ref{app:jhdep} that the Ising energy of such additional UUD/DDU states
on a maple-leaf lattice also remains unchanged in varying $J_{h1}=-J_{h2}$.
%
\section{Ground state}
\label{sec3}
\begin{figure}[tbp]
   \includegraphics[width=0.5\textwidth]{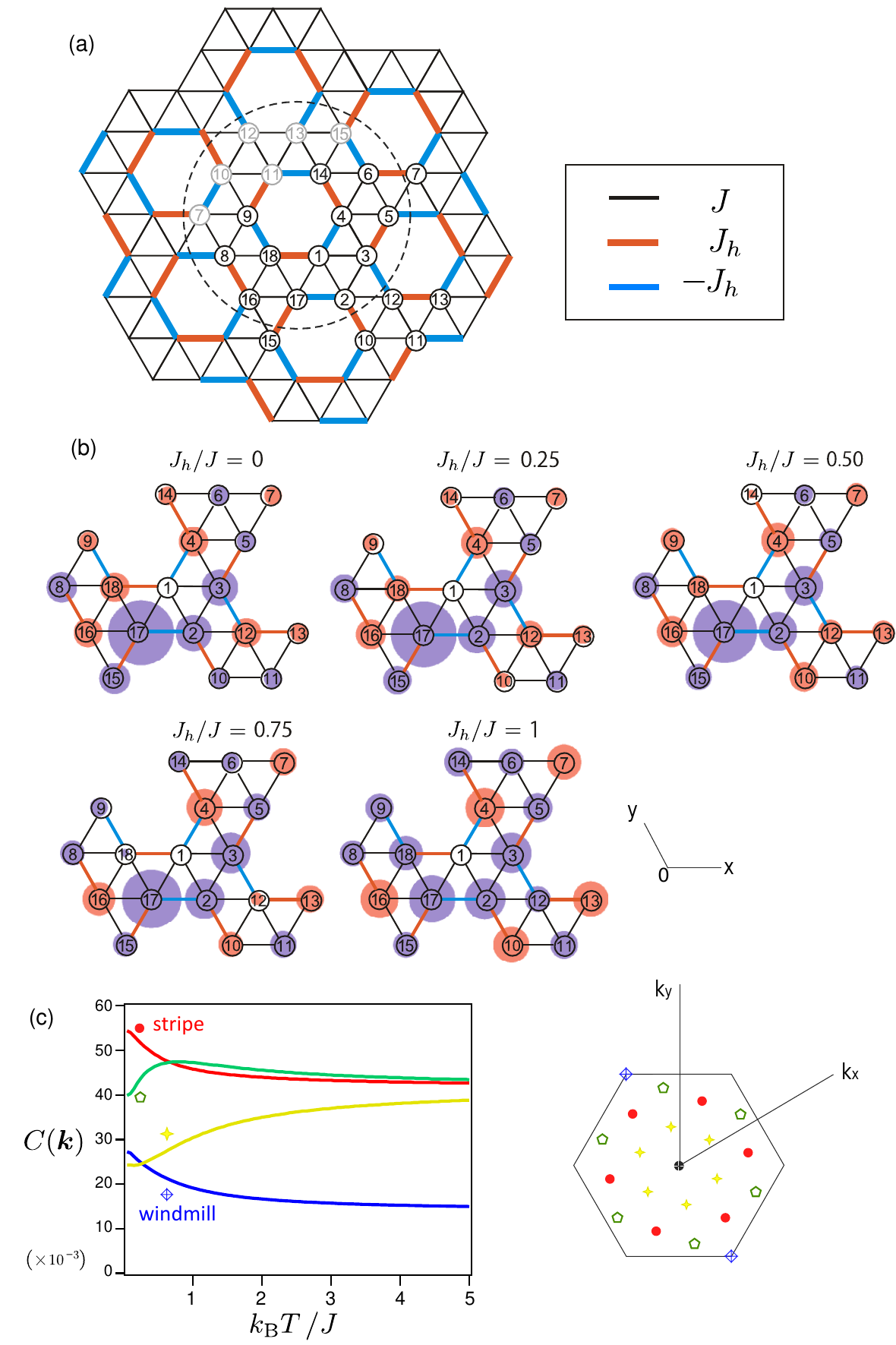}
  \caption{(a) Maple-leaf lattice and 18-site finite-size cluster with periodic boundary conditions.
  (b) Spin-spin correlation functions , $\langle S_{1}S_{j}\rangle$,
   between site 1 and site j of a ground state for several choices of $J_h/J$ obtained by ED.
   The area of each circle is proportional to the magnitude of $\langle S_{1}S_{j}\rangle$,
   with red/purple ones being positive/negative (ferromagnetic/antiferromagnetic).
  (c) Spin structural factor $C(k)$ in Eq.(\ref{eq:st}) calculated for $N=18$ cluster.
   The $k$-points inside the Brillouin zone of the maple-leaf lattice are shown in different symbols
   which are classified into five groups based on $R3$ space group.
   We have $\bm k=(4n-m)/21 (1,-1/\sqrt{3}) + (5n-4m)/21(0,2/\sqrt{3})$ with
   representative data points as $(n,m)=(1,1)$ for stripe (red bullet),
   $(0,\pm 1)$ (yellow star), $(0,\pm 2)$ (green pentagon)
   and $(\pm 1,\mp 3)$ for windmill type (blue diamond).
   The temperature dependences of $C(k)$ is obtained by taking an average of over 50 realizations of
    the TPQ state(see Section \ref{sec4} for details).
}
  \label{f5}
\end{figure}
\begin{figure*}[tbp]
  \includegraphics[width=0.95\textwidth]{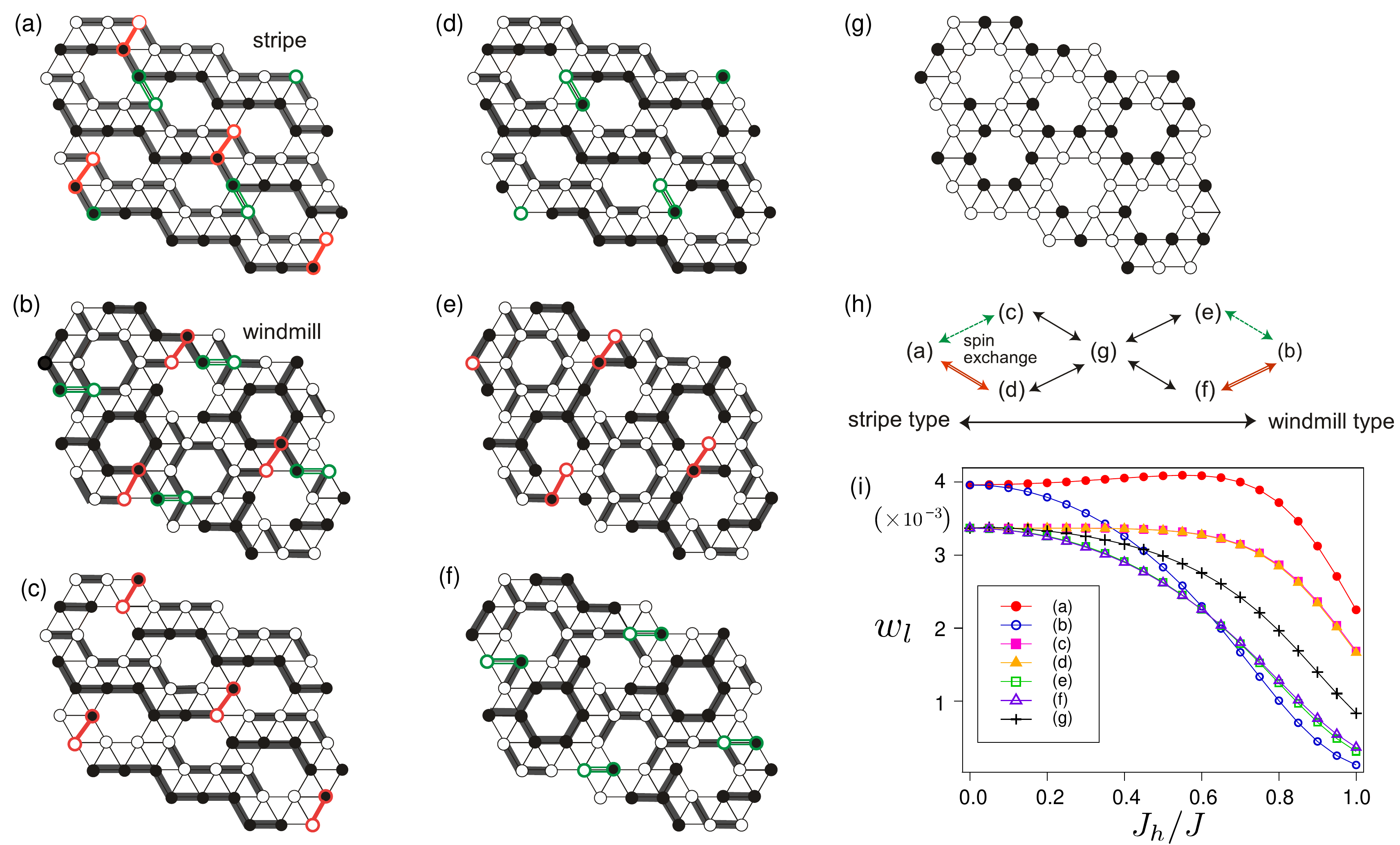}
  \caption{(a)-(g) Representative configuration of spins belonging to UUD/DDU-1 states
   that give major contribution (large $w_l$) to the ground state wave function.
   (a) Stripe and (b) windmill states are the most symmetric ones.
   Panel (h) shows the relationships between states (a)-(g);
   states (c) and (d) can be obtained from (a) by periodically
   exchanging the nearest neighbor up and down spins on green and red bonds, respectively.
   States (e) and (f) can be obtained from (b) in the same manner.
   In panel (i) we plot the actual weights $w_l$ of states (a)-(g) as functions of $J_h/J$
   when $J_{h1}=-J_{h2}=J_h$. At $J_h/J=1$ there appears some non-UUD configurations having
   larger $w_l$ than the (b), (e), (f), and (g) states.
   }
  \label{f6}
\end{figure*}
In this section, we numerically analyze the Heisenberg Hamiltonian Eq.(\ref{Heisenberg})
and show that the unique stripe ground state is possibly formed by the order-by-disorder effect from
the lowest UUD/DDU states of the Ising limit of the maple-leaf lattice we discussed in the previous section.
\subsection{Exact diagonalization}
\label{3-1}
We first perform an exact diagonalization (ED) for $N=18$ clusters
with periodic boundary conditions (see Fig.~\ref{f4}(a)) and obtain the ground state of the maple-leaf lattice.
As we see shortly, the extended magnetic unit cell has 18-sites,
and the larger size available $N=24, 30$ do not match this unit.
This mismatch artificially destabilizes some of the magnetic structures, which we want to avoid.
Choosing $N=18$ enables the full classification of basis states
which is convenient to examine their stability in an unbiased manner.
We additionally examined the $N=24$ ground state in parallel and
confirmed that the major conclusions obtained for $N=18$
do not change which we explain in the final part of this section.
To understand the magnetic property of the ground state we calculate the correlation function
and obtain a structural factor,
\begin{equation}
C(\bm k)= \frac{1}{N} \sum_{i,j=1}^N e^{i r_{ij} k} \langle S_{i}S_{j}\rangle.
\label{eq:st}
\end{equation}
We first show in Fig.~\ref{f5}(b) the correlation functions,
$\langle S_{1}S_{j}\rangle$, in a bubble chart, where the area of the bubble indicates the
strength of the correlation.
A stripe pattern develops along 15-17-(1-)2-3-5-6-15
for all values of $0<J_h/J<1$, and 9-(8-)18-17-2-12-11-9 for $J_h/J\approx 1$.
The correlation between site 1 and site 17 is the most prominent but becomes weaker by increasing $J_h/J$,
and at $J_h >J/2 $ the inequivalence of the magnitude of the correlation between sites becomes smaller.
This tendency is consistent with the Ising energy diagram of Fig.~\ref{f2}(c) where
at $J_h>J/2$ the non-UUD excited states become the lowest.
\par
Figure~\ref{f5}(c) shows the structural factor in Eq.(\ref{eq:st}),
where the $\bm k$-points are the discretized reciprocal points for the $N=18$ cluster.
The $\bm k$-points inside the Brillouin zone can be classified into five groups, depending on the symmetry of the system.
One finds that the peak at $\bm k=(2\pi/7, 10\sqrt{3}\pi/21)$ shown
in red bullet which originates mainly from the stripe-type correlation
dominates the ground state.
This tendency lasts up to $k_BT/J \lesssim 0.5$
(for the details of the calculation, see the next section),
which is the peak position of the susceptibility we see shortly.
The same tendency is observed for other values of $J_{h}/J$.
\par
To understand the underlying mechanism for the development of one-dimensional-like correlation,
we examine the types of basis that have a major contribution to the ground state wave function,
$\ket{\mathrm{GS}}=\sum_{l}c_{l}\ket{l}$.
It is spanned by the total-$S_z=0$ space, $\{\ket{l}\}$ of $l=1,...,\,_N C_{N/2}$,
which are the classical Ising spin configuration we discussed earlier.
Several series of states $\ket{l}$ having large weights $w_{l}=|c_{l}|^2$ are extracted;
we select totally 102 basis states in descending order of $w_l$,
which are classified into seven groups of states.
Their representative configurations are shown in Figs.~\ref{f6}(a)-\ref{f6}(g).
Within each group, the states are related by the $\pm2\pi/3$ rotational and spin inversion symmetries (not shown).
The stripe pattern (a) is one-dimensional and the windmill pattern (b)
is a regular two-dimensional UUD/DDU configuration.
The relationships between these seven groups of states are summarized in Fig.~\ref{f6}(h);
types-(c) and (d) can be obtained from type-(a) by exchanging one nearest neighbor pair of spins
per 18 sites periodically, which are marked by green and red bonds.
Types-(e) and (f) can be obtained from type-(b), and finally
type-(g) with the lowest symmetry can be obtained from types-(c), (d), (e) and (f) by the same process.
In this sense, types-(c) and (d) are similar to (a) and are called stripe-type,
(e) and (f) are similar to (b) and are the windmill-type.
\par
The weights $w_l$ of states (a)-(g) as functions of $J_h/J$ are shown in Fig.~\ref{f6}(i).
The relationships between them are understood as follows:
for $J_h/J=0$, (a) stripe and (b) windmill have the same weight, and (c), (d), (e), (f), and (g) have the same weight as well.
By introducing $J_h$, the stripe (a) and its analogs (c) and (d) dominate the ground state
which continues for $0\leq J_h/J\leq1$.
The second highest contributing configuration is (b) for $0<J_h/J\lesssim 0.37$, but
it is overtaken by the stripe-type (c) and (d) for larger $J_h$.
At $J_h/J=1$, although the stripes (a), (c) and (d) continue to have the largest $w_l$,
the other non-UUD type of spin configurations suddenly become dominant compared to (b) and
other windmill types of states.

In viewing this characteristic feature of the ground state by restarting from the classical Ising limit,
even though we introduce the quantum fluctuation effect,
$J^{xy}_{i,j}$-term in Eq.(\ref{Heisenberg2}),
the major contribution to the ground state is still a series of UUD/DDU states
for all values of $0\le J_h/J \le 1$.
In fact, the states (a)-(g) in Fig.~\ref{f6} all belong to the lowest UUD-manifold of states
called UUD/DDU-1
of Fig.~\ref{f2}(c) with $E_{\rm ising}/J=-5N/24$.
At $J_h/J \ge 1/2$, the other non-UUD states have the lowest Ising energy,
whereas in the Heisenberg model,
the full UUD/DDU states overwhelm the non-UUD states
due to the energy gain from the quantum fluctuations.

\subsection{Energy Gain}
\begin{figure*}[tbp]
  \includegraphics[width=0.9\textwidth]{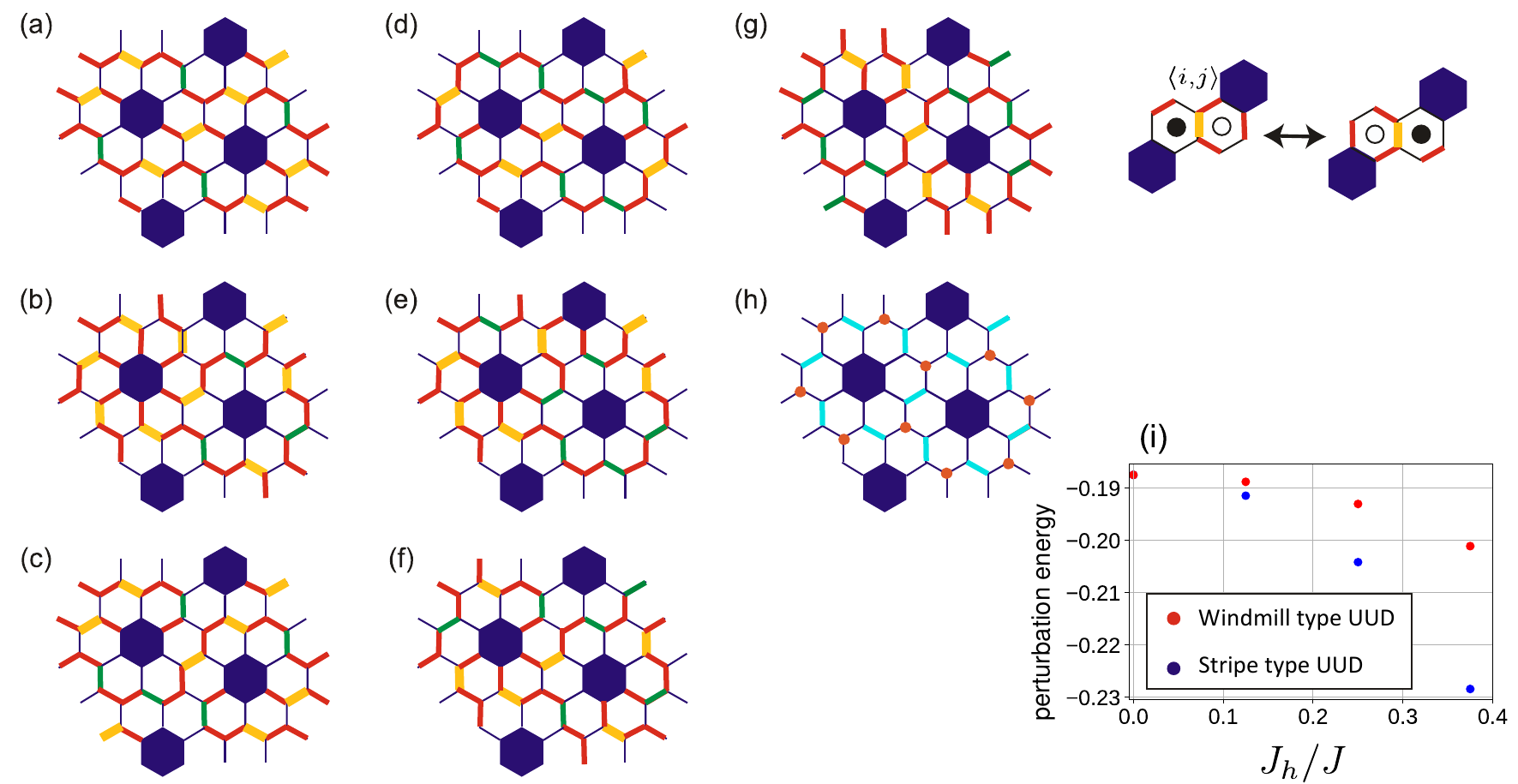}
  \caption{(a)-(g) Bond configurations on the dual lattice corresponding to (a)-(g) in Fig. \ref{f6}.
  Edges shown in yellow/green lines ($J_d$) are always occupied with bonds in these manifolds and
  spin-exchange around a yellow line yields another UUD configuration having the same classical energy
  while that around a green line does not.
 (h) All the yellow/green bonds in (a)-(g) are shown here in the same colored bonds.
  A bond configuration can be determined by identifying bond tilings on the edges connected
  to the vertices (orange dots) in this manifold.
 (i) Energy gain at the second-order of ${\cal H}_{xy}$
  where all the processes of going into other manifolds from UUD/DDU-1 and coming back are considered.
}
  \label{f7}
\end{figure*}

To understand the reason why the stripe ground state is realized,
we examine the effect of quantum fluctuations on the Ising UUD states.
We start from the Ising model and introduce
${\cal H}_{xy}=J^{xy}_{i,j}(S_i^+S_j^-+S_i^-S_j^+)/2$
by setting $J^{xy}_{i,j}$ much smaller than the Ising interaction $J^z_{i,j}$
in Eq.(\ref{Heisenberg2}).
We first rewrite the UUD/DDU configurations in Fig.~\ref{f6} (a)-(g)
using the loop-string model on the dual lattice as shown in Fig.~\ref{f7}(a)-(g).
As mentioned earlier, the UUD/DDU structures correspond to the fully packed states of loops and strings.
Since the length of the loop-string is the number of pairs of neighboring up and down spins,
the fully packed loop-string states are maximally flippable in overall
and may gain $E_q=\langle{\cal H}_{xy}\rangle/N_{\rm cell}$
the most compared to other classical Ising states.
In this context, the ones in Fig.~\ref{f7}(a)-(g) have an equivalent length of loops and may equally contribute to the ground state.
However, in reality, the way how ${\cal H}_{xy}$ works differs between these seven groups of states.
\par
We focus on UUD/DDU-1 and operate one of the ${\cal H}_{xy}$-terms to this manifold.
If we obtain a state that again belongs to UUD/DDU-1,
this term mixes the states within the UUD/DDU-1 manifold at the first order of $J^{xy}_{i,j}$.
If a term in ${\cal H}_{xy}$ transforms the UUD/DDU-1 to the state outside this manifold,
by operating a proper ${\cal H}_{xy}$-term again, we may come back to the UUD/DDU-1 state;
this process serves as a second-order perturbation.
\par
We first consider the first-order perturbation.
The right panel (inset) of Fig.~\ref{f7} shows
the exchange process of antiparallel spins between a pair of nearest neighbor sites $\langle i,j\rangle$
and a pair of hexagons on the dual lattice surrounding the two spins.
The center vertical bond shared by the $\langle i,j\rangle$ hexagons remains unchanged,
whereas occupation of bonds on the other four edges of each hexagon is converted from unoccupied to occupied or vise versa.
This is because if the spin on $i$-th site flips up-side-down,
the ferromagnetic neighbor becomes an antiferromagnetic one and vise versa.

In Figs.~\ref{f7}(a)-(g), we classified the color of occupied bonds that belong to the loop-string
into red, yellow and green.
The yellow and green bonds are the $J_d$-bonds.

The UUD/DDU-1 are characterized as those having all these yellow and green bonds to be occupied.
When we flip the spins on yellow bonds, the number of occupied bonds,
namely the number of red bonds associated with this process remains unchanged,
and the UUD/DDU-1 state stays within the UUD/DDU-1 manifold.
Therefore, we call this yellow bond a flippable bond (see Appendix \ref{app:flip}).
The two structures, stripe-(a) and windmill-(b) have the maximum number of yellow bonds amounting to 2/3 of the flippable bonds
while other structures only have 4/9 of them.
This means that stripe and windmill-states gain the energy the most at the first-order perturbation
by maximally mixing with other UUD/DDU-1 states.

As for the green bond, the spin-exchange will transform the UUD/DDU-1 state to the
non-UUD/DDU states outside the manifold.
In Fig.~\ref{f7}(i), we evaluated the second-order perturbation energy gain
by starting from the windmill or stripe-state, operating ${\cal H}_{xy}$ twice
and summing up the energy gain from each process where we set $J^{xy}_{i,j} = J_{i,j}$ for simplicity.
When $J_h/J>0$, the stripe state becomes more stable than the windmill state.
Flipping the spins on red bond also transforms the UUD/DDU-1 state outside the manifold
and can be treated the same as the green bond.

To be precise, the way how ${\cal H}_{xy}$ mixes the low energy states
is more complicated including not only the UUD/DDU-1 but other UUD/DDU's and the non-UUD/DDU state.
Still, the above discussion gives an overall intuitive understanding
that the stripe-order appears due to the quantum order-by-disorder effect from the classical degenerate
UUD/DDU-1 manifold of states.

We finally mention that the $N=24$ cluster which is larger than the $N=18$ we adopted,
does not accommodate the windmill and stripe type of structures.
However, in calculating the ground state for $N=24$ we find that the overall tendency obtained
for $N=18$ remains unchanged;
the UUD/DDU-1 manifold dominates the ground state,
and the largest contribution to the ground state is the irregular stripe
even though the shape of the stripe is modified due to the mismatch of the shape of the cluster.

\section{Finite Temperature}
\label{sec4}
The ground state of this model possibly breaks the translational symmetry of the original lattice,
and form a stripe type N\'eel order.
To understand the relevance of this ground state with the characteristic behavior
of the magnetic susceptibility resembling those of the one-dimensional Bonner-Fisher curve,
we calculate the finite temperature properties of the model.
Here, we apply a thermal pure quantum (TPQ) method using $N=18, 24$, and 30 clusters.
Similarly to finite-temperature ED methods\cite{jaklic94,aichhorn03,hams2000},
this method gives thermodynamic quantities at finite $N$
by few sample averages\cite{sugiura2012}.
Starting from the high-temperature random state
prepared based on the Haar measure, $|0\rangle=\sum_i C_j |j\rangle$, where $C_j$ is a
random complex coefficient and $\{ \ket{j}\}$ is the Fock space of a finite size lattice,
and successively operating the Hamiltonian $(l-{\cal H}/N)$ shifted by a constant $l$,
a series of states $|k\rangle =(l-{\cal H}/N)^k |0\rangle$ that represent the thermal equilibrium
at different temperatures are obtained.
By using these series of states, we obtain a magnetic susceptibility.
We took more than 30 sampling averages for $N=24$
while for $N=30$ which is the size large enough to represent the thermal state without random average
we used a single calculation.

\subsection{Magnetic susceptibility}
\begin{figure*}[t]
  \includegraphics[width=1.0\textwidth]{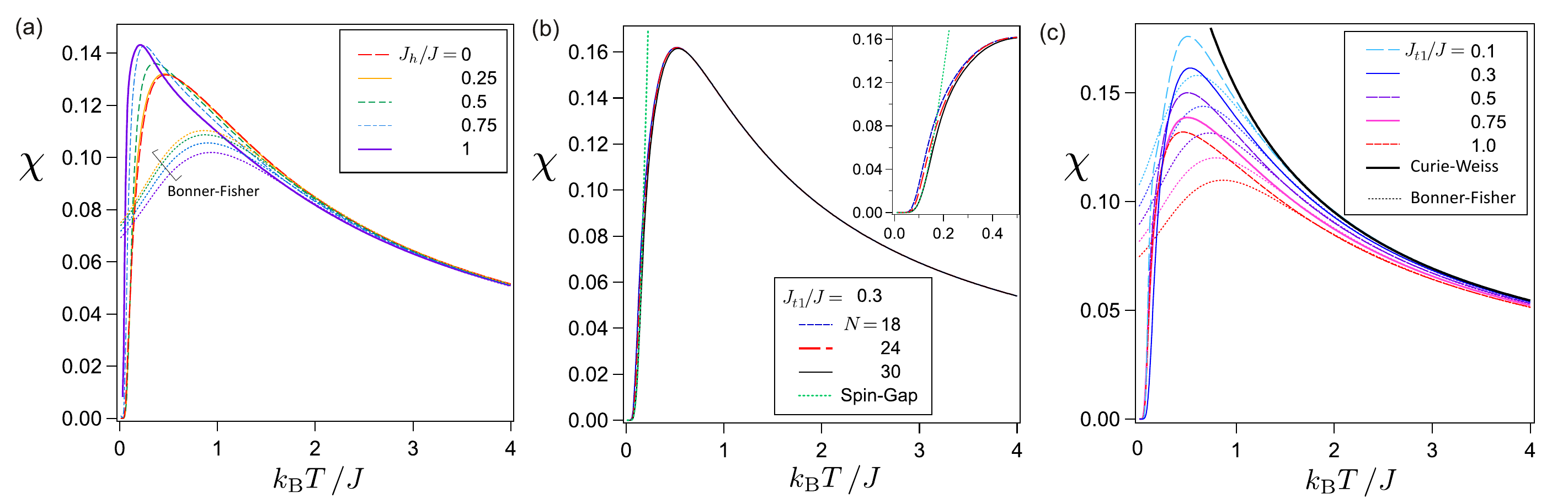}
  \caption{(a) Uniform magnetic susceptibility $\chi$ for several choices of $J_{h1}=-J_{h2}=J_{h}$
   with $J_{t1}=J_{t2}=J_{d}\equiv J=1$.
   (b) $\chi$ obtained for $N=24$ and 30
    with $J_{t2}=J_{d}=1, J_{h1}=-J_{h2}=0.1$ and $J_{t1}=0.3$.
    Inset shows the magnified $\chi$ at low temperature.
   (c) $\chi$ for several choices of $J_{t1}/J$
   with $J_{t2}=J_{d}=1$ and $J_{h1}=-J_{h2}=0.1J$.
   The solid bold line is the Curie-Weiss susceptibility
   $\chi_{cw}=J^{cw}/4k_B(T-\Theta)$ with $J^{cw}=0.928$, $\Theta=-0.396$,
   whose values are obtained by fitting $\chi$ for $J_{t1}=0.1J$.
   Broken lines in both panels (a) and (c) are the Bonner-Fisher susceptibility whose parameter
   $J^{1d}$ is adjusted to fit to the susceptibility of the maple-leaf lattice for each parameter.
}
  \label{f8}
\end{figure*}
First we set $J_{t1}=J_{t2}=J_{d}=J$, $J_{h1}=-J_{h2}=J_{h}$ as we did for the ground state.
A uniform magnetic susceptibility $\chi$ of a $N=24$ cluster for various choices of $J_h$ is
shown in Fig.~\ref{f8}(a).
At low temperature, we find $\chi\propto (\Delta/k_BT)^{1/2}e^{-\Delta/k_BT}$
basically for all choices of $J_h/J$, which indicates a spin-gapped ground state.
For a finite cluster calculation, a fictitious spin gap of order $\Delta \sim 1/\sqrt{N}$
often appears as an artifact of finite size effect.
To examine whether the observed gapped behavior is an intrinsic property of the model,
we compared the result of $N=30$ cluster with $N=18,24$ ones in Fig.~\ref{f8}(b).
The magnitude of spin gap extracted by fitting $\chi$ at low-temperature yields
$\Delta \sim 0.507 (N=24)$ and $0.588 (N=30)$, which increases with $N$.
For the temperatures above the peak position, the difference of $\chi$ between the two system sizes is almost negligible.
From these results, one can judge that the spin gap is finite.

Although several experimental measurements on the maple-leaf materials
suggest that the measured $\chi$ follows a Bonner-Fisher curve characteristic of
a gapless one-dimensional Heisenberg antiferromagnet,
our result with a spin-gapped ground state does not conform to such correspondence.
In Fig.~\ref{f8}(a), we plotted together with
the Bonner-Fisher curve by adjusting its Heisenberg interaction $J^{1d}$
to fit the high temperature ($k_{\mathrm{B}}T>1.5J$) tail of $\chi$ to the ones of the maple-leaf lattice.
For all cases, $\chi$ of the two models do not agree at $k_BT\lesssim J$.
In fact, $\chi$ develops toward lower temperatures than the Bonner-Fisher ones with the higher peaks at lower temperatures.
Such development of peak is the characteristic feature of the two-dimensional highly
frustrated antiferromagnet, e.g. a kagome lattice ones\cite{chisa18}.

To show that this conclusion is not due to our specific choices of parameters,
we examine overall variations of parameters of the model.
Particularly, we investigate the small $J_h/J$ region,  where the two $\chi$'s have relatively better correspondence.
Figure~\ref{f8}(c) shows the case where one of the triangular units $J_{t1}$ is varied while other
parameters are fixed to $J_{t2}=J_{d}=J$, $J_{h1}=-J_{h2}=J_{h}=0.1J$.
Note that the properties of an Ising energy diagram and ground state described in Sec. \ref{sec2} and \ref{sec3}
hold for these choices of parameters.
Again although we properly adjusted $J^{1d}$ for the Bonner-Fisher plot,
the two models do not give a consistent $\chi$ in its amplitude.
However, the peak position becomes closer to each other by decreasing the parameter down to $J_{t1}/J \sim 0.1$
As a reference we also show the Curie-Weiss susceptibility $\chi_{cw}=J^{cw}/4k_B(T-\Theta)$ with $J^{cw}=0.928$, $\Theta=-0.396$
for the same spin density as the maple-leaf lattice to compare with other $\chi$.

\subsection{One dimensional magnon propagation and the XXZ model}
The spin gap and the enhancement of $\chi$ indicate that the low energy effective model
of the maple-leaf lattice is not a uniform antiferromagnetic Heisenberg chain.
However, there is another model, an XXZ model that has a finite spin gap
an enhanced susceptibility, the absence of finite temperature phase transition
and the symmetry-breaking long-range N\'eel ordered ground state.
All these features do not contradict the results we obtained for the present system.

We thus compare the susceptibility of the one-dimensional antiferromagnetic XXZ model,
\begin{equation}
\mathcal{H}_{1d}=\sum_{i}J^{1d}_{xy}(S^{x}_iS^{x}_{i+1}+S^{y}_iS^{y}_{i+1})+J^{1d}_z S^{z}_iS^{z}_{i+1},
\label{xxz}
\end{equation}
where besides $J^{1d}_{xy}$ one can adjust $J^{1d}_z >J^{1d}_{xy}$ to fit the susceptibility as well as the spin gap.
The temperature dependence of $\chi$ is obtained by a size-free calculation using a sine-square deformation
combined with the TPQ method\cite{chisa18} which gives $\chi$ of the thermodynamic limit.

\begin{figure}[tbp]
  \includegraphics[width=0.5\textwidth]{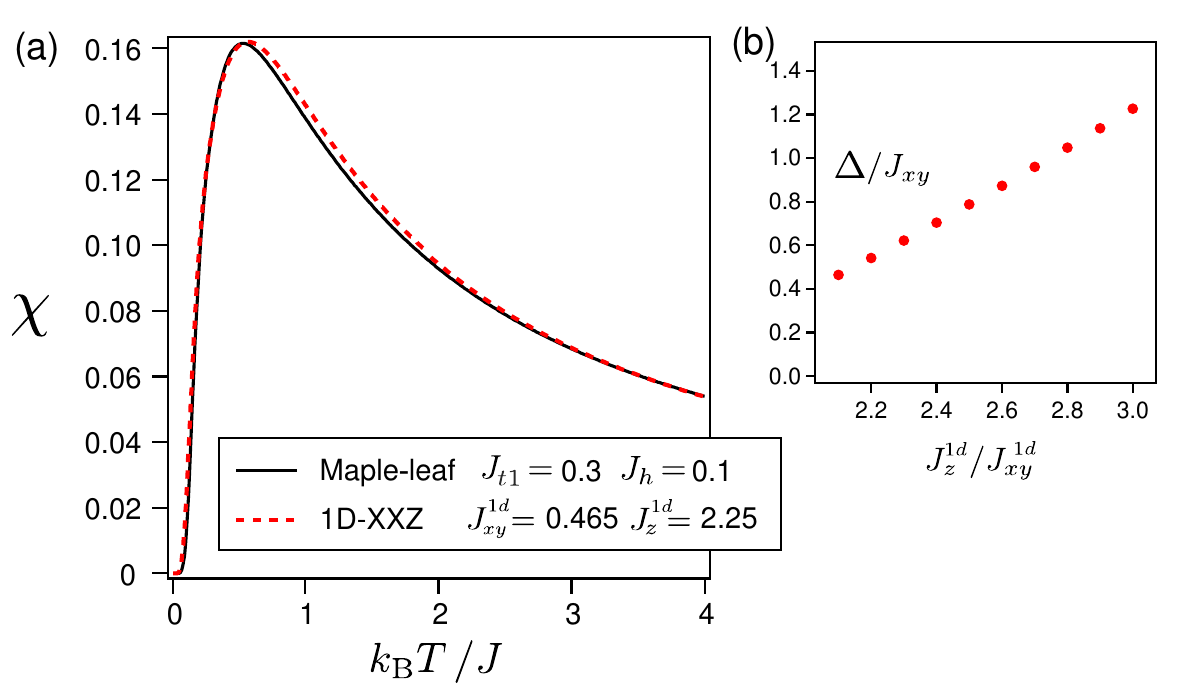}
  \caption{
 Magnetic susceptibility with $J_{t2} = J_d = 1$,
$J_{t1}= 0.3$, and $J_{h1}=-J_{h2}=0.1$ compared with the magnetic susceptibility of XXZ
model with $J^{1d}_{z}/J^{1d}_{xy}=2.25$, where $\Delta=0.54$ .}
  \label{f9}
\end{figure}

As shown in Fig.~\ref{f9}(a), the profile of $\chi$ with smaller $J_{t1}/J$ can be fitted
well with magnetic susceptibility of one-dimensional XXZ model for all temperature regions.
Here, we obtain $J^{1d}_{xy}=0.465$, $J^{1d}_z/J^{1d}_{xy}=2.25$ and
$\Delta=0.581$ for the XXZ model
(see Fig.~\ref{f9}(b)) which is the exact solution of a spin gap of Eq.(\ref{xxz})
that gives the best fitting.
\begin{figure}[tbp]
  \includegraphics[width=0.5\textwidth]{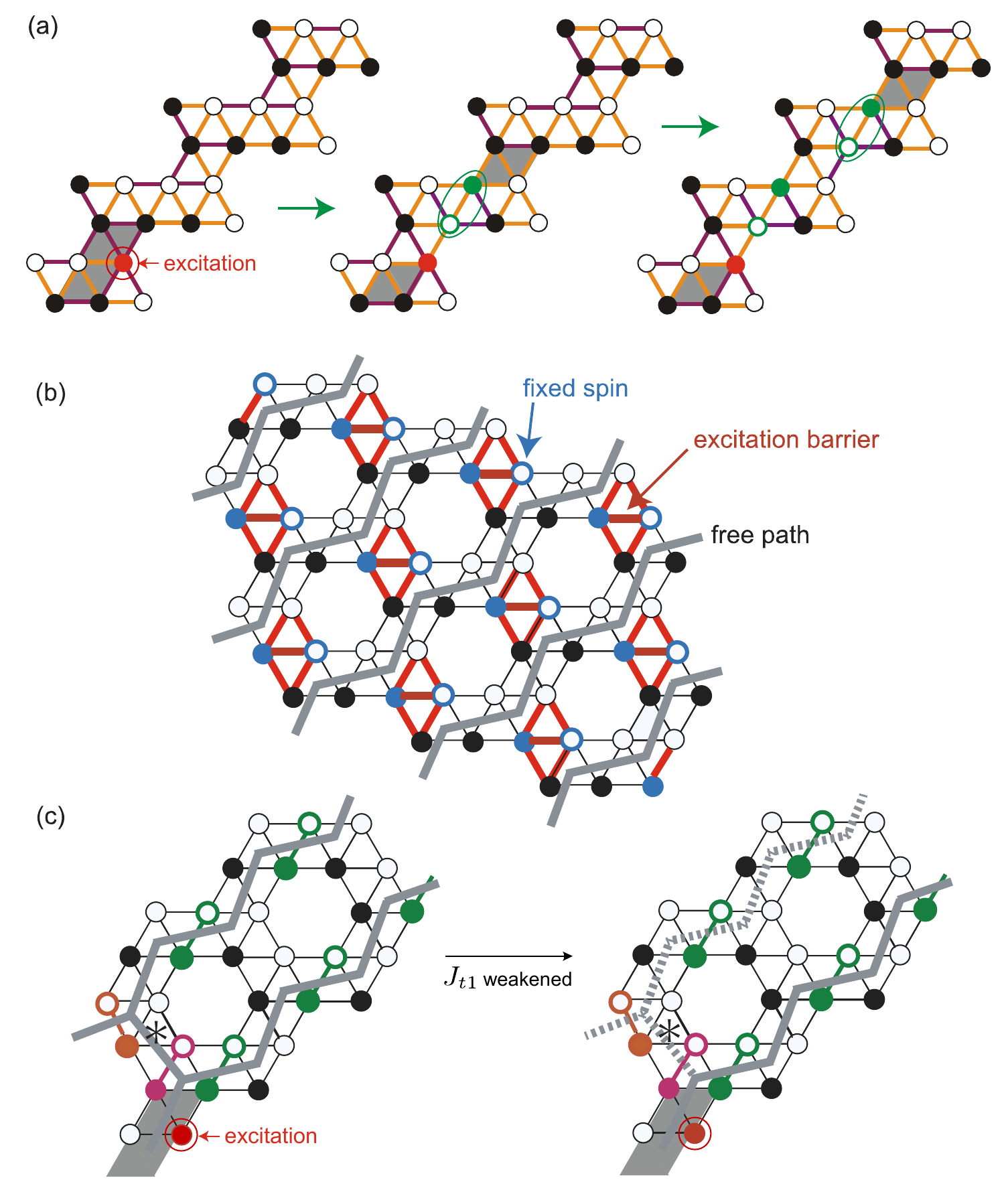}
  \caption{(a) Example of magnon excitation from a stripe.
  A red bullet marked with a circle denotes the location of the excited spin,
  and the adjacent two shaded plaquettes is a magnon that separates into two spinons.
  When ${\cal H}_{xy}$ is operated, the up and down pairs of spins on a green bond marked with oval
  exchanges and a spinon (shaded plaquette) hops to the upper right by one lattice spacings.
  These spin exchanges do not change the Ising energy.
  Bonds gaining/losing energy are shown in orange/purple lines.
  (b) Schematic illustration for the spinon propagation rules.
  Blue open circle indicates the spins that rarely contribute to the spin-exchange process
  (fixed spin) since a finite Ising energy loss occurs.
  Propagation in the direction across the red lines is energetically disadvantageous,
  and the direction that keeps the Ising energy unchanged is restricted to those along the gray lines.
  (c) Path marked with star denotes the exceptional propagation of spinon
  through exchanging orange and red spins, which appears only when the magnon is excited
  on a particular site marked with a circle.
  Weakening $J_{t1}$ from the left panel to the right,
  weakens the connection through starred path and strengthen the one-dimensionality. }
  \label{f10}
\end{figure}
\par
To understand the origin of the coincidence between $\chi$'s of the two models,
we examine a single magnon excitation of the maple-leaf lattice.
For simplicity, we consider a classical stripe UUD/DDU-1 configuration which gives a dominant
contribution to the ground state of the Heisenberg model.
Figure~\ref{f10}(a) shows the example of how the excited magnons propagate
by the spin-exchange process, ${\cal H}_{xy}$.
To identify the location of magnons we use a pair of shaded plaquettes consisting of four triangles
extending in the upper-right direction.
When the system is in the UUD/DDU state such plaquettes
are magnetically neutral (unshaded) since we always have even numbers of up and down spins.
By flipping a down spin to up,
the two plaquettes sharing that site are magnetized as UUUD, which host one magnon.
Unlike the standard Ising models, the spin-exchange
that keeps $E_{\rm ising}$ unchanged takes place, not at the neighboring bonds of
the excited magnon {\it site}, but at the next nearest neighbor,
i.e. it is the neighboring bond of {\it the shaded plaquette}.
By the exchange of two spins marked with ovals in Fig.~\ref{f10}(a),
one of the plaquettes hops to the upper-right by two lattice spacings.
The same operation will propagate the plaquettes in the upper-right or lower-left directions along the stripe.
Since the two-plaquettes share one magnon,
and since this propagation separates two plaquettes freely along the one-dimensional direction
at the first-order level of ${\cal H}_{xy}$,
each shaded plaquette is regarded as spinon.
In the one-dimensional antiferromagnetic XXZ model or Ising model,
a similar spinon excitation is observed
above the spin gap, and this would explain the resemblance of $\chi$ between the two models.
\par
The one-dimensionality of this spinon propagation is confirmed as follows.
There is a relatively larger energy cost of flipping the spins
marked with blue-open symbol in Fig.~\ref{f10}(b);
its Ising energy gain with its surroundings
amount to $-(J_d+2J_{t1})/4$ or $-(J_d+2J_{t2})/4$,
larger than for the other spins $-J_d/4$ that join the spinon propagation.
Accordingly, the exchange of spins marked with solid and open blue circles
have a large energy loss of
$(J_{t1}+J_{t2})/2$, and work as a spatial barrier.
There are two barriers per every depleted hexagon,
which restrict the propagation of magnons
to a one-dimensional direction in the gray line.

There is an exceptional case that may slightly allow the two-dimensional propagation;
in Fig.~\ref{f10}(c) the path marked with the star had an energy barrier in the stripe UUD state
but once a magnon is excited on a particular site indicated by an arrow,
the energy barrier is lost and the magnon can propagate
by exchanging spins on the orange and pink bonds.
However, the energy barrier is lost only for the limited choices of excitation,
and the overall nature of the magnon propagation is regarded as one-dimensional.
A smaller $J_{t1}/J$ weakens the connection through the starred path
because $J_{t1}$ is responsible for the energy gain due to the exchange of spins on the orange bond.
Consequently, the smaller $J_{t1}/J$ strengthens the one-dimensionality of the propagation pathway.
This may explain the reason why our magnetic susceptibility with smaller $J_{t1}$
gives better correspondence to the ones for the one-dimensional XXZ model.
\par

As can be understood from the similarities of the nature of the magnetic excitations,
the dimensional reduction is expected only for the magnetic properties of the system.
We examined the specific heat of the two models for the same parameters in Appendix \ref{app:c_mh},
and found that the nonmagnetic part of the low energy excitations differs between the two models.
Whereas, the onset of the magnetization curve of the maple-leaf lattice (see Appendix \ref{app:c_mh})
shows a criticality reminiscent of the gapped 1D spin system,
which is another sign of the dimensional reduction effect.
The parameter range that the dimensional reduction is observed is limited to
the case of small $J_{h1}=-J_{h2}$.
For all antiferromagnetic $J_{h1}=J_{h2}>0$, the system recovers a two-dimensional N\'eel ordering
(see Appendix \ref{app:phase}).

\subsection{Three dimensionality due to inter-layer coupling}
\begin{figure}
  \includegraphics[width=0.4\textwidth]{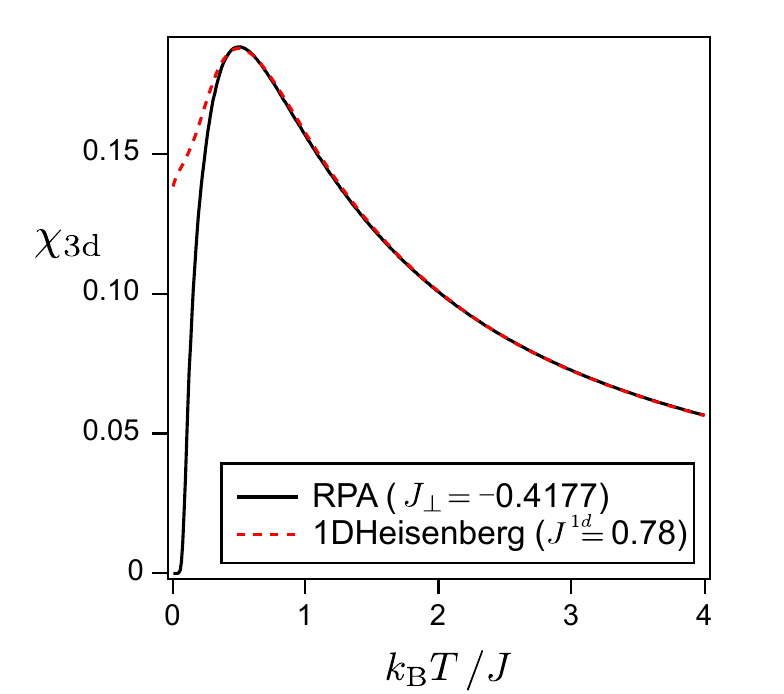}
  \caption{Magnetic susceptibility $\chi_{\rm 3d}$
with $J_{t2}=J_{d}=1, J_{h1}=-J_{h2}=0.1$ and $J_{t1}=0.30$
considering interplanar interaction $J'=-0.4177$ with RPA.
Broken line is the Bonner-Fisher curve of a one-dimensional Heisenberg model.}
  \label{f11}
\end{figure}
In the previous subsection, we figured out that the susceptibility of the maple-leaf lattice
can be fitted well throughout the whole temperature range
by the one-dimensional XXZ model with appropriate choices of the ratio of $J_z/J_{xy}$ as
well as $J_{xy}$ that also reproduces a finite spin gap of the ground state.
Compared to the Bonner-Fisher curve for the one-dimensional Heisenberg model,
these susceptibilities show large enhancement toward low temperature with higher peaks.

On the other hand, the experimental observation indicates that for bluebellite
the susceptibility shows almost perfect coincidence with the Bonner-Fisher curve down to temperatures
just below the peak position\cite{Bluebellite}.
In further lowering the temperature, there occurs a phase transition in experiments to the N\'eel ordered state
due to the three-dimensionality of the material,
which masks the intrinsic low-temperature property of the purely two-dimensional magnet.

Here, we examine whether our XXZ-like susceptibility may reproduce the Bonner-Fisher curve except
at temperatures lower than the spin gap.
We take account of a layered structure of the maple-leaf lattices stacked in the $z$-direction
and deal with the interplanar interaction $J_\perp$ using random phase approximation (RPA).
The three-dimensional susceptibility $\chi_{3\mathrm{d}}$ described using
the intra-layer $\chi_{2\mathrm{d}}$ is given as
\begin{equation}
\chi_{3\mathrm{d}} = \frac{\chi_{2\mathrm{d}}}{1+2J_\perp\chi_{2\mathrm{d}}}.
\end{equation}
We adopted several choices of inter-layer interaction $J_\perp$.
As shown in Fig.~\ref{f11}, for a ferromagnetic interplanar interaction $J_\perp<0$
using $\chi_{2\mathrm{d}}$ of the maple-leaf lattice with $J_{t1}<0.3$,
$\chi_{3\mathrm{d}}$ shows good agreement with the Bonner-Fisher curve at $0.5\lesssim k_BT/J$,
including the peak.
This shall be because a strong magnetic fluctuation characteristic of the frustrated lattice at low temperature
is suppressed by the three dimensionalities.
Such agreement can be found for parameters $J_t/J\lesssim 0.3$ with a proper choice of $J_\perp$.

\section{Summary and Discussion}
\label{sec:summary}
We examined the ground state and finite temperature magnetic properties of the spin-1/2 maple-leaf lattice,
which is the 1/7-depleted triangular lattice with spatially modified interaction strength.
Although the maple-leaf lattice structure with five independent exchange interactions
may seem rather complicated, the intrinsic nature of the model can be understood
by considering the Ising model and examining the nature of the low energy states.
We find that the degree of frustration is seemingly weakened from the triangular lattice
by the depletion; the originally highly degenerate UUD/DDU states
are divided into few manifolds of states with relatively smaller degeneracies.
However, this degeneracy is still large enough to contribute to a finite residual entropy.
By varying the exchange interactions systematically, we examined the degree of frustration
measured by the constraint parameter.
The frustration is not much different from that of the triangular lattice
and may become comparably strong depending on the choices of parameters.
\par
The ground state of the maple-leaf lattice Heisenberg model turned out to be
a possibly translational-symmetry-broken stripe state.
This state emerges from order-by-disorder.
Starting from the lowest UUD/DDU manifold of states in the Ising limit,
the quantum fluctuations in the spin-exchange term mix them.
Based on the analysis of spin patterns,
we found that the stripe-pattern UUD/DDU state gains the quantum fluctuation energy the most
and is selected as a ground state.
\par
When a single magnon is excited from the stripe ground state,
it splits into two spinons, each propagating along the one-dimensional path formed by the stripe.
Because of the one-dimensional alignment of spins,
there arises an energy barrier between the sinks (depleted sites),
which hinders the spinons to hop to the other neighboring one-dimensional path.
The presence of a spin gap indicates that the Ising energy loss
of exciting a single magnon is larger than the kinetic energy gain from such restricted motion of spinons.
This kind of spinon excitation is very similar to the spin-1/2
one-dimensional antiferromagnetic XXZ model with a large spin gap.
The magnetic susceptibility of a maple-leaf lattice can be well fitted with the
susceptibility of this XXZ antiferromagnet.
\par
Since the original maple-leaf lattice is a two-dimensional system,
the observed one-dimensional magnetic property is regarded as a dimensional reduction phenomenon.
Similar dimensional reduction is observed previously in an anisotropic triangular lattice
where one bond direction among the three has a stronger interaction than the other two;
for a Heisenberg antiferromagnet, the low energy spectrum is composed of an incoherent continuum indicating
the spinon-like propagation\cite{Kohno2007}.
It explained the spectrum of the inelastic neutron scattering in Cs$_2$CuCl$_4$\cite{Coldea2003}.
Also for spinless fermions, the fractionalization of excited charges is observed,
which is the analog of spinons of magnets and is recently referred to as ``fractons". The fractionalized excitation gives a
similar continuum in the one-particle excited spectrum\cite{Hotta2008}.
For a wider class of materials, the stacked layered magnets often have
inter-layer interactions that connect one site with more than two.
These frustrated interactions cancel out and the inter-layer correlation is suppressed,
which is observed in the critical exponent of BaCuSi$_2$O$_6$ near the quantum critical
point\cite{Sebastian2006,Roesch2007}.
In a cubic antiferromagnet called Pharmacosiderite\cite{Okuma2021} based on an octahedron,
each inter-octahedra interaction consists of six bonds which are frustrated, and
the reduction of dimensionality from three- to two- and even to one-dimension is observed
in neutron experiments.
If the inter-layer interactions in three dimensions and the inter-chain interaction in two-dimensions
are practically weaker than the intra-layer or intra-chain ones,
the dimensional reduction is encoded in the system. The frustration among weak inter-layer/chain interactions plays a secondary role.
\par
In that context, the intrinsic difference of our maple-leaf lattice from most of the above examples
except Pharmacosiderite is that the original lattice structure and the Hamiltonian is
spatially isotropic and purely two-dimensional.
The parameter region we find the phenomena is restricted to a small but finite range of
$J_{h1}=-J_{h2}$ and $J_{t1}$ (see Appendix \ref{app:phase} for details),
namely the bond interaction $J_{ij}$ is not spatially uniform.
However, the system retains the symmetry of the original lattice and is invariant
under the $C_3$-rotation, and there is no reason to favor a one-dimensionality
in the geometry of $J_{ij}$ itself.
The dimensional reduction of the magnetic excitation occurs spontaneously due to a strong frustration effect.
\par
We finally presented the scenario that the experimentally observed Bonner-Fisher-like susceptibility
may not necessarily be a coincidence.
The enhanced susceptibility due to the strong fluctuation is characteristic of the frustrated magnetism.
By introducing a three-dimensionality, namely the inter-layer magnetic exchange interaction,
the enhancement can be suppressed, and the peak height becomes closer to the Bonner-Fisher curve.
Experimentally, only the temperature regions down to slightly below the peak was the target range
that the Bonner-Fisher fitting functioned.
The peak temperature is typically $k_BT\lesssim J$ and since $\chi$ behaves closer to the Curie-Weiss-type
ones at $k_BT \gtrsim 2J$, the characteristic feature of magnetism manifests only
in the peak temperature, peak height, and the behavior below the peak.
Therefore, it is more likely that even though the dimensional reduction indeed takes place in the
material, the recovery of three-dimensionality at temperatures below $k_BT \lesssim J$ due to weak
but finite inter-layer coupling will push $\chi$ to a more realistic phase.
Below that temperature the phase transitions induced by the three-dimensionality is
observed in bluebellite.

To clarify more intrinsically of how the nature frustration changes of depletion remains a future issue.
The present study shows that the depleted series of frustrated magnets can be
a good platform to interpolate many different types of frustrated lattices and fill the
black parameter space to be examined in theories as well as in experiments.

\section{acknowledgement}
We thank Yuya Haraguchi and Zenji Hiroi for the discussions.
This work was supported by JSPS KAKENHI Grants No. JP17K05533,
JP18H01173, No. 20K03773, No. JP21H05191, No. JP21K03440
from the Ministry of Education, Science, Sports and Culture of Japan.
The calculations were partially performed using the Supercomputer Center, the Institute for Solid State Physics, the University of Tokyo.

\appendix
\section{General proof of $J_h$-independence of UUD/DDU states when $J_{h1}=-J_{h2}=J_h$}
\label{app:jhdep}
\begin{figure}[tbp]
  \includegraphics[width=0.3\textwidth]{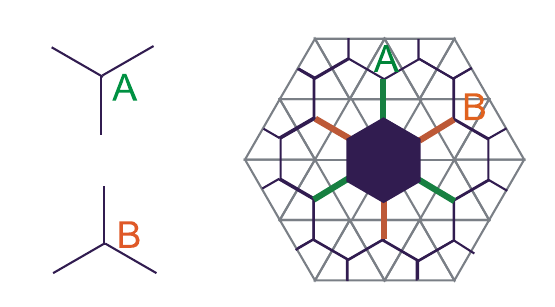}
  \caption{Label of the dual lattice by A and B. All edges going into sinks from $J_h$ sides is connected to A vertices and those going into sinks from $-J_h$ is connected B vertices}
  \label{f12}
\end{figure}
\begin{figure}[tbp]
  \includegraphics[width=0.5\textwidth]{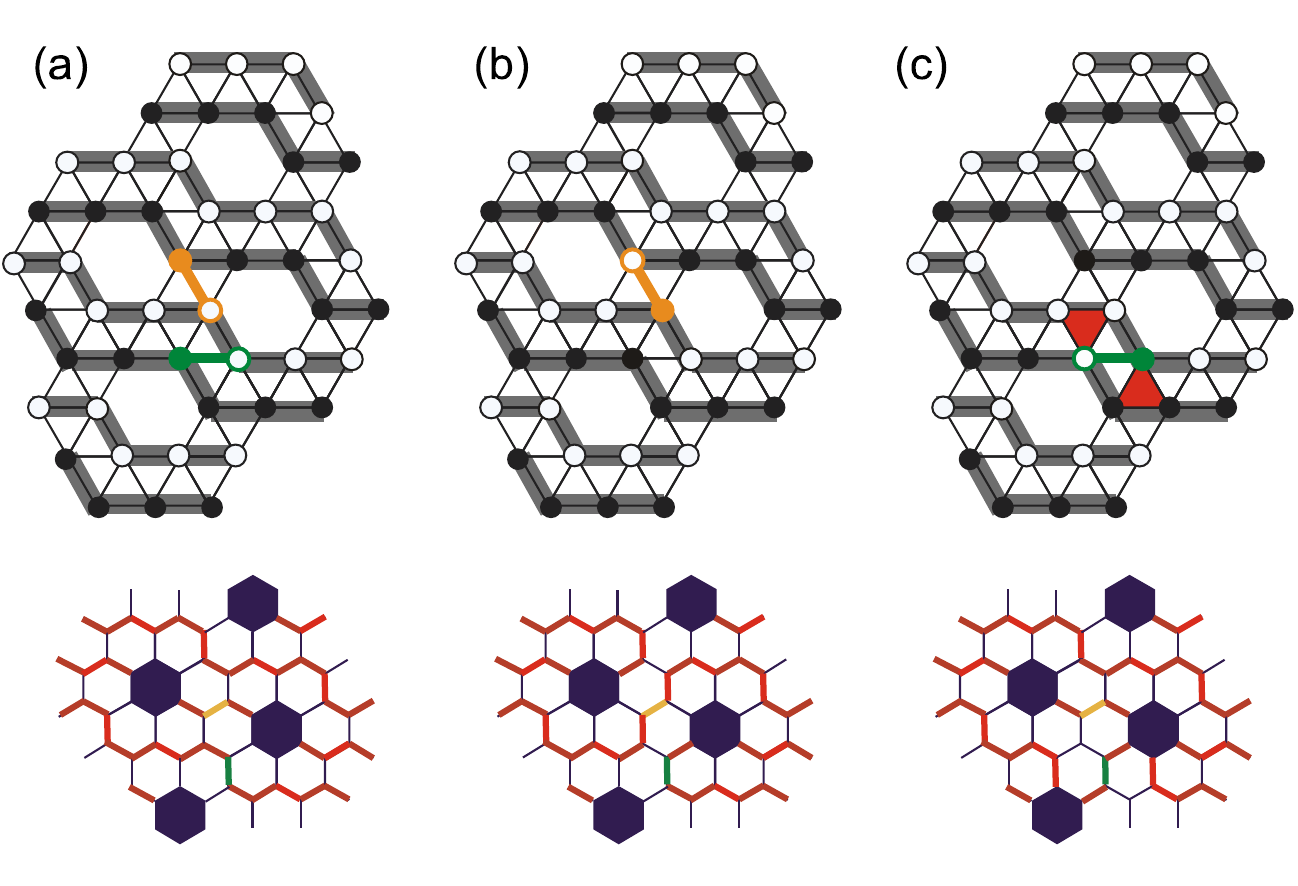}
  \caption{Example of configuration of spins on a maple-lattice
  and loop-strings on a honeycomb lattice with sinks in (a) UUD-1 stripe state,
 (b) when flipping a yellow bond from panel (a), and (c) flipping a green bond from panel (a). }
  \label{f13}
\end{figure}

In this section, we give proof to support the discussion in Sec.\ref{sec3}
that the energy values of all UUD/DDU states are independent of $J_h$ in the case of $J_{h1}=-J_{h2}=J_h$.
Notice that all UUD/DDU states in the maple-leaf lattice are mapped one-to-one to a fully packed
loop-string configuration on its dual lattice.
Then, one only needs to prove in the fully packed loop-string state
that the number of loop-bonds on a dual lattice going into a sink crossing the $J_h$-edge of the hexagon of the maple-leaf lattice
is equal to the number of loop-bonds crossing the $(-J_h)$-edge.
Let us label the vertices on the dual lattice by ``A'' and ``B'' as shown in Fig.~\ref{f12}.

In a fully packed loop-string state, the number of bonds connected to A vertices is equal to that to B vertices
since the number of bonds connected to each vertex is the same between the two.
This means that the number of bonds going into a sink from A vertices and that from B vertices
are the same; this is because the bond not going into a sink always connects an A vertice and a B vertice
and does not affect the balance between the numbers of bonds connected to A and B.
Since all edges going into sinks through $J_h$ is connected to A vertices and those going into sinks
through $-J_h$ is connected B vertices, our statement is proved.

\section{Effect of ${\cal H}_{xy}$ on the UUD-1 stripe state}
\label{app:flip}
We show in Fig.~\ref{f13} how the spin configuration of UUD-1 stripe in panel (a)
changes by flipping the yellow bond to the other UUD-1 state in panel (b).
When flipping the green bond in panel (a) the non-UUD state in panel (c) is realized.
The corresponding loop-string configuration on the honeycomb lattice with sink is shown
in the lower panels.
Notice that the UUD/DDU triangles are not equivalent in their Ising energy unlike the triangular lattice antiferromagnet.
This is because the UUD structure does not necessarily optimize the energy of each triangle locally;
there are positive and negative $\pm J_h$ which can be smaller in amplitude than $J$.

\section{Comparison of magnetic susceptibilities of several models}
\label{app:sus}
\begin{figure}[t]
  \includegraphics[width=0.4\textwidth]{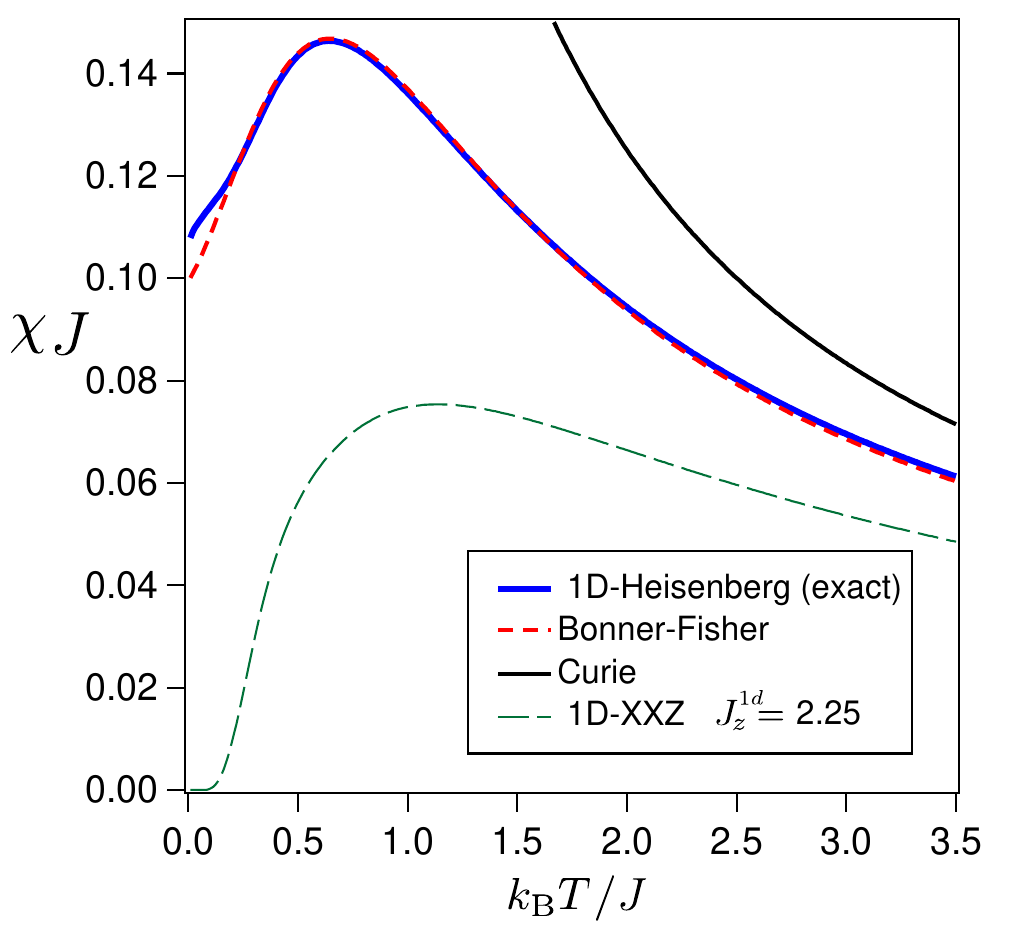}
  \caption{Comparison of magnetic susceptibility of one-dimensional antiferromagnetic Heisenberg model,
  Bonner-Fisher plot, one-dimensional antiferromagnetic XXZ model for several choices of
  $J^{1d}_{z}/J^{1d}_{xy}$ with $J^{1d}_{xy}=J$, and Curie plot $J/4k_{\mathrm{B}}T$.}
  \label{f14}
\end{figure}
The comparison of magnetic susceptibilities of several models is shown in Fig.~\ref{f14}.
The magnetic susceptibility of the one-dimensional Heisenberg model is obtained
from exact solution using quantum transfer matrix method\cite{Klumper00} and
its result is numerically evaluated within a negligibly small error.
The Bonner-Fisher curve\cite{Bonner1964} which is first determined by the finite-size scaling to the finite cluster ED results obeys
\[
\chi_{\mathrm{BF}} = \frac{0.25+0.14995 x+0.30094 x^2}{k_{\mathrm{B}}T
(1+1.9862 x+0.68854x^2+6.0626x^3)},
\]
with $x=J^{1d}/2k_{\mathrm{B}}T$.
One finds that the Bonner-Fisher curve deviates from the exact solution at $k_BT/J\lesssim 0.4$.
The logarithmic drop of $\chi$ takes place at extremely low temperature below the data obtained,
which requires a more accurate numerical integration of Ref.[\onlinecite{Klumper00}].
The Curie plot is given by $\chi_{\mathrm{C}}=1/4k_{\mathrm{B}}T$.
The magnetic susceptibility of the one-dimensional XXZ model
is obtained for $J^{1d}_z/J^{1d}_{xy}=2.25$(the same as in Fig.\ref{f9})
by a size-free calculation using a sine-square deformation combined
with the TPQ method\cite{chisa18}.

\begin{figure}[t]
  \includegraphics[width=0.5\textwidth]{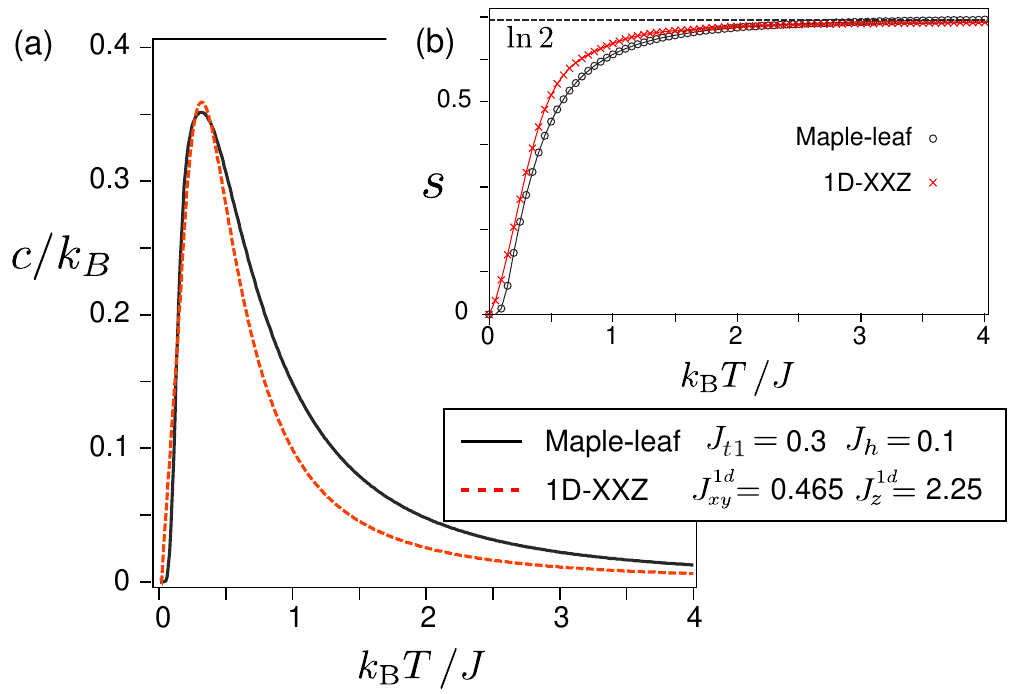}
  \caption{(a) Specific heat per site $c/k_{\rm B}$ calculated using TPQ and sine-square deformation. The parameters of the maple-leaf Heisenberg model and 1D XXZ model are the same as Fig.~\ref{f9}.
  (b) Entropy density for the data in panel (a). }
  \label{f15}
\end{figure}

\section{Other physical quantities}
\label{app:c_mh}
The dimensional reduction effect appears in the magnetic properties of the system.
To clarify this point, we show in Fig.~\ref{f15}(a) the specific heat per site $c/k_B$ of the maple-leaf lattice
Heisenberg model with $J_{t1}=0.3$ and $J_h=0.1$ which are the same parameters with Fig.~\ref{f9}.
Here, we compare it with the result of the 1D XXZ model.
For the 1D XXZ model, we used the TPQ combined with a sine-square deformation\cite{chisa18} and extracted the center energy bond
to evaluate the energy. The specific heat is obtained by the derivative of the energy.
The $N=20$ and 24 results agree within the accuracy of ${\cal O}(10^{-3})$
with basically negligible size effect characteristic of the sine-square deformation\cite{chisa12}.
The horizontal axis is scaled as $J^{1d}_{xy}=0.465$ in the same manner as Fig.~\ref{f9};
the peak positions of the $c/k_B$ of two models agree very well, while above the peak temperature, there are extra contributions for the
maple leaf lattice.
\par
The difference may come from the nonmagnetic contribution to the specific heat.
The nonmagnetic lowest energy excitation of the XXZ model exists below the spin gap,
but for the maple-leaf lattice, the lowest energy excitation is the magnetic one.
As we saw in Fig.~\ref{f5}(c), the stripe ordering is weakened above the peak temperature,
and the other UUD/DDU-1 and non-UUD states may appear. These extra contributions would explain the difference.
The entropy obtained by integrating $c/T$ is given in Fig.~\ref{f15}(b) which
supports that below the spin gap $\sim 0.5 J$ the nonmagnetic contributions appear only for the XXZ model.
\par
\begin{figure*}[tbp]
  \includegraphics[width=0.95\textwidth]{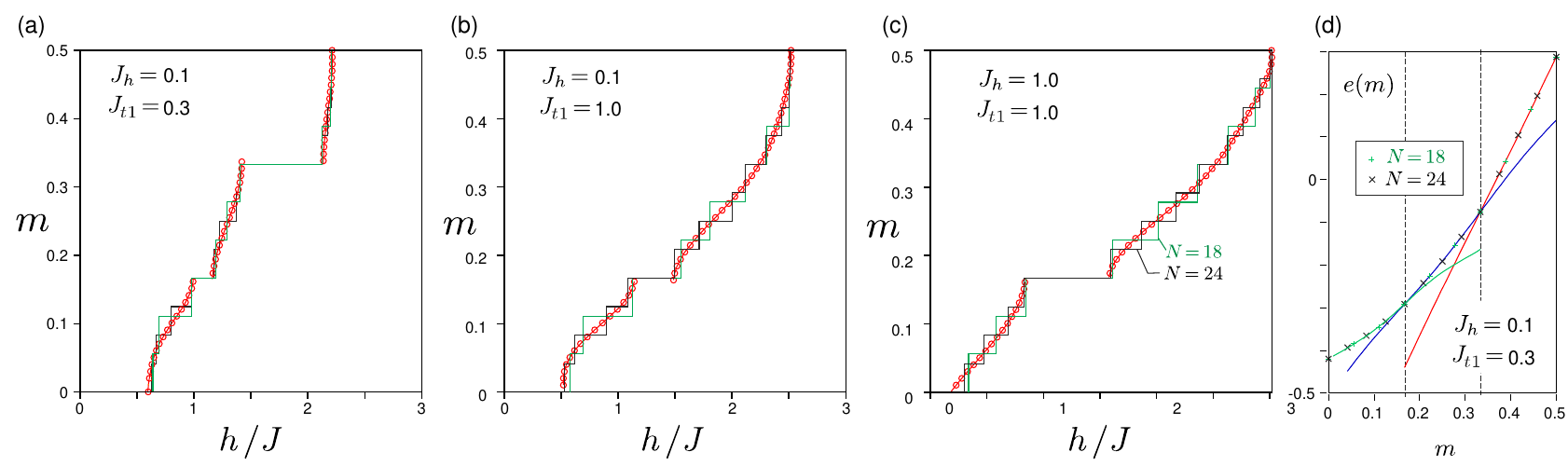}
  \caption{
  Magnetization curve obtained for (a) $(J_h,J_{t1})=(0.1,0.3)$,
  (b)$(0.1,1.0)$ and (c)$(1.0,1.0)$.
  The stepwise data in solid lines are from the exact diagonalization on $N=18$ and 24 clusters and
  the data points on curves are obtained using the kernel regression method.
  Panel (d) shows the energy $e(m)$ of $N=18$ and 24 corresponding to panel (a)
  as a function of magnetization density $m=\langle s^z\rangle/N$,
  and the three lines are the results of the kernel regression given independently for
  regions separated by the 1/3 and 2/3 plateaus.
  We take the mirror for the data below and above the plateaus following Ref.\onlinecite{nakamura20}.
  }
  \label{f16}
\end{figure*}

Figure~\ref{f16} shows the magnetization curve of the system, where we plotted the stepwise structure
obtained by the exact diagonalization for $N=18$ and 24 clusters
and the curve obtained by applying a kernel regression method\cite{nakamura20,harada11} to these data.
Here, we chose the data below and above the 1/3-plateaus separately.
For $J_{t1}=0.3$ we also have 2/3-plateau for $J_h=0\sim 1$, which we applied the same treatment.
Using the discrete energy levels $e(m)$ from ED as a function of magnetization density $m$,
a continuous function $e(m)$ is obtained for each region.
Since the value $h=\partial e(m) / \partial m$ at $m=1/6$ differs for those obtained based on the data at $m<1/6$ and $m>1/6$
we find the 1/3-plateau.
For $J_h/J=0.1$ and $J_{t1}=0.3$ the onset of the curve reminds us of a square-root critical behavior,
$m\propto |h-h_c|^{1/\delta}$ with $\delta=2$ characteristic of a one dimensional spin gapped system\cite{sakai1998}.
However, for $J_h/J=1$ it becomes close to $\delta\sim 1$ expected for two-dimensional quantum magnets\cite{nakano2011}.
These results also support the dimensional reduction of the system observed for small $J_h/J$.

\begin{figure}[tbp]
  \includegraphics[width=0.5\textwidth]{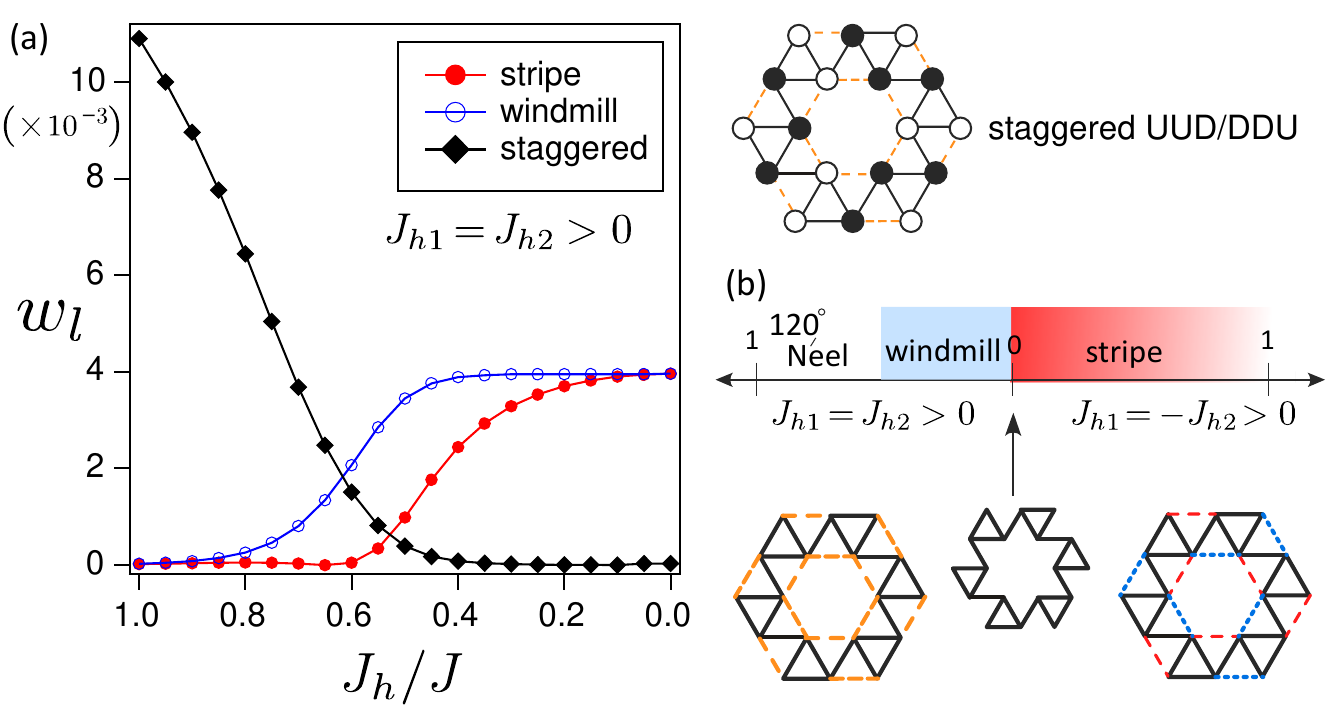}
  \caption{(a) Weights $w_l$ for three major contributions
   for the fully antiferromagnetic $J_{h1}=J_{h2}=J_h$ are shown.
   The staggered state has the antferromagnetic correlation along the hexagon,
   and appears only for this fully antiferromagnetic case.
   (b) The overall phase diagram obtained by varying $J_{h1}=\pm J_{h2}$. }
  \label{f17}
\end{figure}

\section{Antiferromagnetic case $J_{h1}=J_{h2}>0$ }
\label{app:phase}
We examine the case where all the interactions are antiferromagnetic.
Figure~\ref{f17}(a) shows the weight of the ground state wave function $w_l$ for $N=18$.
The stripe, windmill and staggered type configurations show the major contribution.
The parameter range corresponds to Fig.~\ref{f2}(b).
The staggered state shown in the right panel is the one not found for $J_{h_1}=-J_{h_2}$ case,
and show the antiferromagnetic correlation around the hexagon.
This state and the windmill state are spatially isotropic with a purely two-dimensional character,
and the staggered state contribute to the $120^\circ$ N\'eel order reported earlier\cite{Schulenburg2000,Schmalfuss2002}.
The overall phase diagram is shown in Fig.~\ref{f17}(b).
The stripe and the dimensional reduction are expected for small and mixed $J_{h_1}=-J_{h_2}$ region.

\bibliography{ref}
\end{document}